\documentclass[aps,prd,reprint,onecolumn,notitlepage,superscriptaddress]{revtex4-1}

\usepackage{lipsum}
\usepackage{amsmath}
\usepackage{breqn}
\usepackage{graphicx}
\usepackage{color}
\usepackage{tikz}
\usepackage{verbatim}
\usepackage{hyperref}
\usepackage{amssymb}
\usepackage{bm}
\newcommand{\rom}[1]{\textup{\uppercase\expandafter{\romannumeral#1}}}

\begin{document}

\title{Scalar and vector modes in inflation with antisymmetric tensor field}
\author{Abhijith Ajith}
\email{abhijith.ajith.421997@gmail.com}
\affiliation{Indian Institute of Science Education and Research$,$ Bhopal$,$ $462066,$ India}
\author{Sukanta Panda}
\email{sukanta@iiserb.ac.in}
\affiliation{Indian Institute of Science Education and Research$,$ Bhopal$,$ $462066,$ India}

\begin{abstract}
We investigate the scalar and vector modes arising from cosmological perturbations within the framework of an inflationary scenario driven by an antisymmetric tensor field, minimally coupled to gravity. After eliminating gauge artifacts, there remain four scalar and six vector modes of interest which can be studied separately. We analyze the stability of these modes, while looking for generic instabilities like ghost and gradient instabilities that could potentially plague the theory. Further, we investigate the evolution of these modes across different regimes, particularly subhorizon and superhorizon scales. 

\end{abstract}

\maketitle

\section{Introduction}
The inflationary paradigm attempts to explore our universe in quite an appealing way and tries to explain how our universe came to the way it is.  Historically, several problems such as the horizon problem, the flatness problem, the monopole problem, and so on, which plagued the standard big bang cosmology, were tackled by the idea of inflation. Further, inflation could provide a natural mechanism by associating the idea of quantum fluctuations to explain the formation of large-scale structures that we see in our universe \cite{Riotto:2002yw,Enqvist:2019jkb,Battye:1998xe}. In light of the above, inflationary model building has been of strong interest in the research community. Inflation is generally modeled by either using external fields \cite{Abedi:2016sks,Kodama:2021yrm,Wands:2007bd,Gong:2006zp,Ohashi:2011na,Vazquez:2018qdg,Bartolo:2021wpt}, or by incorporating modifications to gravity \cite{Starobinsky:1980te,Inagaki:2019hmm,Zhang:2021ppy,Sangtawee:2021mhz,Bamba:2015uma,Bhattacharjee:2021kar}. Various cosmological observations put tight constraints on the physics of inflation by probing its imprints on the early Universe. The Cosmic Microwave Background (CMB), from experiments like Planck \cite{Planck:2018vyg,Planck:2019nip,Planck:2018lbu, SPT-3G:2021eoc, ACT:2020frw, WMAP:2012nax, Planck:2018nkj,Lemos:2023rdh,Tristram:2007zz}, constrains inflationary parameters such as the scalar spectral index ($n_s$) and the tensor-to-scalar ratio ($r$). Meanwhile, B-mode polarization measurements \cite{SPT:2019nip,SPTpol:2013omd,Bartolo:2019eac} and observations from large-scale structure \cite{DAmico:2022gki,Rezaie:2023lvi, Chaussidon:2024qni}, further test predictions like primordial gravitational waves and non Gaussianities. Together, these probes rule out or put stringent constraints on many of the existing inflationary models \cite{Iacconi:2019vgc,Planck:2018jri,Planck:2019kim}. Further, additional constraints arising from the swampland conjectures in string theory have further tightened the plight of these models \cite{Andriot:2018mav,Brennan:2017rbf,Obied:2018sgi,Garg:2018reu,kinney2019,Kallosh:2019axr}. This has given room for exploration of other extended models, like vector and tensor field models, that can replicate the inflationary scenario.

Vector field models suffer from generic instabilities like ghost and gradient instabilities \cite{Himmetoglu:2009qi,BeltranJimenez:2017cbn,Golovnev:2011yc}. However, latter studies employing multiple vector fields have shown improvements over their predecessors \cite{Gorji:2020vnh,Murata:2021vnb,jiro2009,jiro2010a}.
Speaking of tensor fields, there exist several studies on inflation driven by non-symmetric tensors, particularly the antisymmetric tensor \cite{Prokopec:2005fb,Koivisto:2009sd,Obata:2018ilf,Elizalde:2018rmz}. The presence of antisymmetric tensor in the early universe is motivated by superstring models \cite{Rohm:1986,Ghezelbash:2009gf,jiro2010b}. Inflationary cosmology employing $n-$forms is discussed in  \cite{jiro2010b,jiro2013,jiro2015,2012JCAP...12..016M,DeFelice:2012jt}. 
 $n-$forms having non-minimal couplings with gravity are capable of mimicing slow roll inflation. The detailed analysis of the background dynamics for the generic n-form inflation model along with their perturbative evolution is studied in \cite{Germani:2009iq,Koivisto:2009sd}. At linear level, the inflationary models with 2-form fields encounter instabilities similar to that of vector field models. For the 3-form field, inflation can be achieved without the requirement of slow roll and can be freed from instabilities \cite{Koivisto:2009sd,DeFelice:2012jt}. Antisymmetric tensor fields of rank 2 and 3 acting as a standalone driving field for inflation was studied in \cite{Koivisto:2009ew,Koivisto:2009sd}, where generic instabilities were highlighted. Later, it was shown that models with single and multiple rank-2 antisymmetric tensor fields can give rise to inflationary solutions in the presence of minimal and non-minimal couplings with gravity \cite{Ajith:2022wia,Aashish:2018lhv,Aashish:2019zsy,Aashish:2020mlw}. Fortunately, these models are devoid of the generic ghost and gradient instabilities, unlike the vector field and 2-form inflation models studied earlier. Further, they predict a nearly scale invariant power spectrum for the tensor modes in quasi de-Sitter limit \cite{Ajith:2022wia,Aashish:2019zsy,Aashish:2021gdf}.

The past studies on inflation employing rank 2 antisymmetric tensor fields have mainly dealt with the background cosmology. The analysis of the perturbations that can arise in these models has not been explored much. The study of cosmological perturbations is of utmost importance, since they play a central role in connecting theoretical models to observable quantities. Cosmological perturbations can be decomposed into scalar, vector, and tensor modes that evolve independently at the linear level \cite{Dodelson:2003ft,Guzzetti:2016mkm}. In our previous studies, we have talked in detail about the tensor modes that are expected to manifest themselves in the form of primordial gravitational waves \cite{Ajith:2022wia,Aashish:2021gdf}. In the case of scalar and vector modes, both the metric and the driving tensor field can source them. The analysis performed in our previous studies is preliminary, as only the individual effects of the scalar and vector modes that come from the driving tensor field have been analyzed so far. A much deeper analysis should involve the contributions from the metric as well as the driving tensor field simultaneously. This is a much heavier task, as we have many modes to handle altogether. In addition, the inherent complex nature of our model further makes their analysis more troublesome. 

In this work, we try to analyze the nature of scalar and vector modes in the context of inflation driven by a single antisymmetric tensor field. The focus is primarily on the nature and evolution of the modes, with comprehensive stability analysis to check the presence of ghost and gradient instabilities, thereby assessing their physical viability.
The organization of this paper is as follows. In section \ref{sec:2} we give a brief overview of the model we analyzed in our previous studies. We talk about the gauge transformations in the context of cosmological perturbations in section \ref{gt}. In section \ref{sec:3}, we look at the scalar modes in our model, emphasizing on the stability conditions and their evolutionary behaviors. In section \ref{sec:4}, we do a similar analysis for the plausible vector modes. We conclude our findings and address the future possibilities in section \ref{sec:5}.

\section{Formalism and background}\label{sec:2}
In this section, we review the general background analysis, carried out for the antisymmetric tensor field inflation model given in \cite{Aashish:2021gdf}. Initially, it was shown that slow roll inflation is plausible with a rank-2 antisymmetric tensor field minimally coupled to gravity. Later, a non-minimal coupling of the field with the Ricci scalar was introduced to the minimally coupled tensor field action to tune the velocity of primordial gravitational waves appearing from the model. The tensor field action looks like,
\begin{equation} \label{eq:1}
     S=\int d^{4}x\sqrt{-g}\left[\frac{R}{2\kappa}-\frac{1}{12}H_{\lambda\mu\nu}H^{\lambda\mu\nu}+\frac{\tau}{2}(\nabla_{\lambda}B^{\lambda\nu})(\nabla_{\mu}{B^{\mu}}_{\nu})-B_{\mu\nu}B^{\mu\nu}(\frac{m^2}{4}-\frac{\xi}{2\kappa}R)-\frac{m^2\phi_0^2}{4}\right]
\end{equation}
Here, $R$ denotes the Ricci scalar and $g$ is the metric determinant. $\kappa$ is the inverse of squared Planck mass ($M_{\rm Pl}^2$) and $\tau$ is a dimensionless parameter. $H_{\lambda\mu\nu}$ is a three ranked tensor called the field strength, defined as, $H_{\lambda\mu\nu}=\nabla_{\lambda}B_{\mu\nu}+\nabla_{\nu}B_{\lambda\mu}+\nabla_{\mu}B_{\nu\lambda}$. This along with the term that contains the parameter $\tau$ constitute the kinetic terms in $B_{\mu\nu}$ for our action \cite{Altschul:2009ae}. Further, out of these terms, the one that contains the field strength tensor represents the gauge invariant kinetic term, whereas the second term with the $\tau$ parameter violates the gauge symmetry. This gauge violating term rescued the  system by eliminating ghost and gradient instabilities, while considering standalone perturbations in $B_{\mu\nu}$. We have considered a quadratic potential in $B_{\mu\nu}$ which has the form $\frac{m^2}{4}(B_{\mu\nu}B^{\mu\nu}+\phi_{0}^{2})$. The background structure of $B_{\mu\nu}$ considered in \cite{Aashish:2021gdf} is such that the quantity $B_{\mu\nu}B^{\mu\nu}$ is negative. Hence, $\phi_{0}^{2}$ is a positive shift added with the intention of making the potential positive. Assuming an isotropic and homogeneous early universe, it is advisable to work with the flat FLRW metric with the components,
\begin{equation}\label{eq:2}
     g_{00}=-1, \hspace{.5cm}  g_{ij}=a(t)^{2}\delta_{ij}
 \end{equation}
 The background field $B_{\mu\nu}$ is structured as,
 \begin{equation}\label{eq:3} 
    B_{\mu\nu}=a(t)\phi(t)\left ( \begin{array}{cccc}
    0 & 1 & 1 & 1 \\
    -1 & 0 & 0 & 0 \\
    -1 & 0 & 0 & 0 \\
    -1 & 0 & 0 & 0 \\
    \end{array} \right )
\end{equation}
The temporal dependence of the scale factor $a(t)$, the Hubble parameter $H(t)$ and the background field $\phi(t)$ is to be assumed from now on until and unless stated otherwise. With this choice, we can see that the equations of motion appear similar to that of a scalar field model. The dynamics of the system is determined by the Einstein equations along with a constraint equation for the conservation of stress energy tensor. Using them, we obtain two independent equations given as,
\begin{equation}\label{eq:4} 
    H^2=\kappa m^2\phi^2-\kappa\tau(H\phi\dot{\phi}+2H^2\phi^2)+\frac{\kappa m^2\phi_0^2}{12}-6\xi(3\dot{H}\phi^2-2H\phi\dot{\phi}+5H^2\phi^2)
\end{equation}
\begin{equation}\label{eq:5}
    2\dot{H}+3H^2=3\kappa\tau(H\phi\dot{\phi}+2H^2\phi^2)+\frac{\kappa m^2\phi_0^2}{4}+6\xi(2\phi\ddot{\phi}+2\dot{\phi}^2+4H\phi\dot{\phi}-\dot{H}\phi^2-3H^2\phi^2)
\end{equation}
An exact de-Sitter inflation where $\phi$ and $H$ remain constant in time can be obtained by constraining the coupling parameter $\xi$. But we cannot necessarily demand a pure de-Sitter case, since it may be possible for $H$ and $\phi$ to have very slight temporal dependencies. This scenario referred to as the quasi de-Sitter or the slow roll inflation, can be realized through the introduction of slow roll parameters. The slow roll parameters in our theory are $\epsilon\equiv -\dot{H}/H^2$ and $\delta\equiv\dot{\phi}/H\phi$ which incorporate the time derivatives of $H$ and $\phi$ respectively. Further, $\delta$ is related to the number of e-folds of inflation. For a viable inflationary model, it should be able to support at least around 70 e-folds of inflation. With this condition it can be seen that $\delta<\frac{1}{70}\ln\left(\frac{\phi_f}{\phi_i}\right)$, \cite{Aashish:2018lhv}. To begin with, the slow roll parameters are considered as small quantities, and hence will be able to satisfy the requirement of 70 e-folds.

Apart from analysing the background dynamics, a preliminary study on the cosmological perturbations arising from this model has already been performed in \cite{Aashish:2021gdf}. The perturbations can arise simultaneously from both the metric side and the  driving field $B_{\mu\nu}$. Following the SVT decomposition, \cite{Dodelson:2003ft,Guzzetti:2016mkm}, the perturbations to the metric tensor can always be expressed in the following form,
\begin{equation}\label{eq:6}
\begin{array}{c}
    \delta g_{00}=-\mathcal{E} \hspace{1cm} \delta g_{0i}=a(\partial_i \mathcal{F} +\mathcal{G}_i) \\
    \delta g_{ij}=a^2(\mathcal{A}\delta_{ij}+\partial_i\partial_j\mathcal{B}+(\partial_i\mathcal{C}_j+\partial_j\mathcal{C}_i)+\mathcal{D}_{ij})
    \end{array}
\end{equation}
where the perturbations $\mathcal{A}$, $\mathcal{B}$, $\mathcal{C}_i$, $\mathcal{D}_{ij}$, $\mathcal{E}$, $\mathcal{F}$, and $\mathcal{G}_i$ are functions of $\Vec{x}$ and $t$, satisfying the conditions,
\begin{equation}\label{eq:7}
    \partial^i\mathcal{C}_i=\partial^i\mathcal{G}_i=0, \hspace{2em} \mathcal{D}_{ij}=\mathcal{D}_{ji} \hspace{2em} \partial^i\mathcal{D}_{ij}=0, \hspace{2em} {\mathcal{D}^i}_i=0
\end{equation}
Thus we can see that $\mathcal{A}, \mathcal{B}, \mathcal{E}$ and $\mathcal{F}$ are the scalar modes, $\mathcal{C}_i$ and $\mathcal{G}_i$ the divergence free vector modes and $\mathcal{D}_{ij}$ constitutes the traceless, transverse tensor modes. The perturbations from the tensor mode $B_{\mu\nu}$ take the following form \cite{Aashish:2018lhv},
\begin{equation}\label{eq:8} 
    \delta B_{0i}=-E_i \hspace{2em} \delta B_{ij}=\epsilon_{ijk}\ \delta^{kl} M_l
\end{equation}
where $\epsilon_{ijk}$ is the completely antisymmetric Levicivita tensor of rank 3. By Helmholtz theorem, we can always decompose a vector into a curl free part and a divergence free part. Applying this to the above vector fields, we obtain,
\begin{equation}\label{eq:9}
    \vec{E}=\vec{\nabla}u+\vec{U}, \hspace{1cm} \vec{M}=\vec{\nabla}v+\vec{V}
\end{equation}
Here $\Vec{U}$ and $\Vec{V}$ are the divergence free vector modes, and $u$ and $v$ constitute the scalar modes. By the decomposition theorem, in a homogeneous and isotropic background, the scalar, tensor, and vector modes evolve separately at linear order and can be studied individually. The preliminary study performed in \cite{Aashish:2021gdf} considered only the tensor part of the perturbations coming from the metric. The absence of tensor modes in the $B_{\mu\nu}$ perturbations helped in studying the metric tensor modes separately since they do not couple with the scalar and tensor parts. In the case of scalar and vector modes, a preliminary analysis was performed while switching off those that come from the metric. While looking at the standalone perturbations that come from $B_{\mu\nu}$, it was observed that generic instabilities like ghost and gradient instabilities could be bypassed if we are to set the parameter $\tau > 0$ \cite{Aashish:2019zsy,Aashish:2020mlw,Aashish:2021gdf}. The tensor modes from the metric were seen to propagate in the form of primordial gravitational waves. Estimation of the primordial GW velocity in the minimally coupled scenario gave us a value roughly equal to $c/\sqrt{3}$, where $c$ is the velocity of light in vacuum. The possibility of such a value for the GW velocity cannot be entirely ruled out, though the observation from neutron star merger data GW170817 has shown that GWs propagate with the velocity of light in vacuum \cite{Odintsov:2019clh}. However, with the addition of a non-minimal coupling term that involves Ricci scalar, the primordial GW velocity could be tuned to $c$, consistent with the recent data on astrophysical GWs.
As of now, the scalar and vector modes arising from the metric have been neglected in the study of perturbations. A better analysis on the perturbations demands the knowledge of these modes as well. Hence, we intend to study the scalar and vector modes in the context of our tensor field model. 
\section{Gauge transformations}\label{gt}
Before addressing the scalar and vector modes in the model, we need to be clear about the issue of gauge in our theory. We are considering the perturbations around a homogeneous and isotropic background. Since general relativity is a diffeomorphism-invariant theory, the perturbations are defined with reference to a chosen coordinate system. Hence, they are not uniquely specified. This leads to the gauge problem. Specifically, infinitesimal coordinate transformations of the form:
\begin{equation}\label{eq4.8}
    x^\mu \rightarrow \tilde{x}^\mu = x^\mu + \xi^\mu(x)
\end{equation}
induce changes in the perturbation variables that may not correspond to any physical alteration of the space-time. As a result, not all perturbations will be physically meaningful, with some of them being pure gauge artifacts.
Under the transformation in eq. (\ref{eq4.8}), the metric tensor transforms as, 
\begin{equation}
\tilde{g}_{\mu\nu}(\tilde{x}) = g_{\lambda\kappa}(x) \frac{\partial x^\lambda}{\partial \tilde{x}^\mu} \frac{\partial x^\kappa}{\partial \tilde{x}^\nu}.
\end{equation}
However, instead of explicitly transforming all the coordinates and fields, we can adopt a passive point of view, by attributing all the changes in the metric to a transformation of the metric perturbation \( \delta g_{\mu\nu} \). The gauge transformation then affects only the perturbation in the following manner \cite{Weinberg:2008zzc}
\begin{equation}
\delta g_{\mu\nu}(x) \rightarrow \delta g_{\mu\nu}(x) + \Delta\delta g_{\mu\nu}(x)
\end{equation}
where $\Delta\delta g_{\mu\nu}(x) \equiv \tilde{g}_{\mu\nu}(x) - g_{\mu\nu}(x)$. The field equations should be invariant under this transformation. To first order in both \( \delta g_{\mu\nu} \) and \( \xi^\mu(x) \), this change is given by the Lie derivative of the background metric along the  field \( \xi^\mu \). This can be explicitly written as,
\begin{equation}
   \Delta\delta g_{\mu\nu} =  - \xi^{\rho}_{,\mu} \, \bar{g}_{\rho\nu} - \xi^{\sigma}_{,\nu} \, \bar{g}_{\mu\sigma} - \bar{g}_{\mu\nu,\alpha} \, \xi^\alpha
\end{equation}
Further the spacial part of the four vector field $\xi^{\mu}\equiv(\xi^0,\xi^j)$ can be decomposed into the gradient of a spatial scalar plus a
divergenceless vector. Applying the expression for the metric perturbation we chose in eq. (\ref{eq:6}), we can write how these decomposed modes transform under the above gauge transformation.
\begin{equation}
    \begin{aligned}
        \Delta \mathcal{A} &= 2H \, \xi^0 \\
\Delta \mathcal{B} &= -\frac{2}{a^2} \, \xi^S \\
\Delta \mathcal{C}_i &= -\frac{1}{a^2} \, \xi^V_i \\
\Delta \mathcal{D}_{ij} &= 0 \\
\Delta \mathcal{E} &= 2\dot{\xi}^0 \\
\Delta \mathcal{F} &= \frac{1}{a} \left( -\xi^0 - \dot{\xi}^S + 2H \, \xi^S \right) \\
\Delta \mathcal{G}_i &= \frac{1}{a} \left( - \dot{\xi}^V_i + 2H \, \xi^V_i \right)
    \end{aligned}
\end{equation}
Here $\xi^S$ and $\xi^V_i$ represent the corresponding scalar and divergence free vector part of the spatial component of $\xi_{\mu}$. Analogously the tensor field perturbations $\delta B_{\mu\nu}$ transform as,
\begin{equation}
   \Delta\delta B_{\mu\nu}= - \xi^{\rho}_{,\mu} \, \bar{B}_{\rho\nu} - \xi^{\sigma}_{,\nu} \, \bar{B}_{\mu\sigma} - \bar{B}_{\mu\nu,\alpha} \, \xi^{\alpha}
\end{equation}
where $\bar{B}_{\mu\nu}$ represents the background value of the antisymmetric tensor field. Hence, the transformation of the modes can be written as,
\begin{equation}\label{eq4.14}
\begin{aligned}
    \Delta\partial_i u &= a\phi_i\dot{\xi}^0+a\phi_j \partial_i\partial^j \xi^S + (\dot{a}\phi_i+a\dot{\phi}_i)\xi^0 \\
    \epsilon_{ijk}\delta^{lk}\Delta\partial_l \ v &=-a\phi_j\partial_i\xi^0+a\phi_i\partial_j\xi^0 \\
    \Delta U_i &= -\frac{1}{a}\phi_j\ \delta^{kj}\ \partial_i\xi_k^V \\
    \Delta V_i &= 0
\end{aligned}   
\end{equation} 
with $\phi_i = \phi(t) $. The tensor mode $\mathcal{D}_{ij}$ is already gauge invariant and no gauge fixing is required for the same. We have utilized this convenience while working with the tensor modes. But the analysis of scalar and vector modes requires the removal of these gauge artifacts. We can eliminate the gauge degrees of freedom either by choosing a gauge or via writing our equations completely in the form of gauge invariant variables. In our analysis that follows, we will be using the former method, i.e analyzing our equations in a particular gauge.
\section{Scalar modes}\label{sec:3}
In this section, we look at the nature of the probable scalar modes in our theory. Scalar modes are of significant interest in inflationary models. Generic scalar field inflation models predict these modes to couple with density perturbations that lead to the growth of the large scale structure we see in our universe. In our model, we have in total six scalar modes of interest, of which four of them originate from the metric while the remaining two are contributed by the $B_{\mu\nu}$ field. Owing to the not-so-simple structure of the tensor field action, these modes are expected to couple with each other, making our analysis difficult. It is convenient to work with the Fourier components of the perturbations. As long as we treat the perturbations as infinitesmal quantities, there is no coupling between the Fourier components of different wave numbers. Further, we can conveniently eliminate the spatial derivatives, $\partial_j$, by replacing them with $-ik_j$. Since we are in the preliminary stage of our analysis, we are employing a simplification by choosing the $Z$ axis to lie along the direction of 3-momentum $\vec{k}$. Thus, for a generic scalar mode $f$, we can write,
\begin{equation}\label{eq:10} 
    f(\vec{x},t)=\int d^3k e^{-ikz}f_{\text{ft}}(\vec{k},t)
\end{equation}
Hence the terms which contain partial derivatives in $x$ and $y$ directions will now go to zero. For notational convenience, we are omitting the 'ft' subscript in the fourier transforms. From now, the usual symbols for the modes will denote their fourier counterparts, unless stated otherwise. For this momentum vector $\vec{k}$, the vanishing divergences of the vector modes make their fourier components in the Z direction vanish, i.e $\mathcal{C}_z$, $\mathcal{G}_z$, $U_z$ and $V_z$ are all 0. Further, the gauge transformations for the modes from $B_{\mu\nu}$, take the form,
\begin{equation}
\begin{aligned}
    -\frac{ik}{a\phi}\Delta u &= \dot{\xi}^0 +\left(H+\frac{\dot{\phi}}{\phi} \right)\xi^0-\frac{k^2}{a^2}\xi^S \\
    \Delta v &= 0 \\
    \Delta U_i &=0 \\
    \Delta V_i &=0
\end{aligned} 
\end{equation}
We can see that the modes $v$, $U_i$ and $V_i$ are independent of the gauge artifacts for this choice. For fixing the gauge, we choose  $\xi^S$ so that $\mathcal{B} = 0$, and then choose $\xi^0$ so that $\mathcal{F} = 0$. This gauge is called the Newtonian gauge \cite{Ma:1995ey}. Both choices are unique, so that after choosing Newtonian gauge, there is no remaining freedom to make gauge transformations. Further, it is conventional to write the scalar modes $\mathcal{E}$ and $\mathcal{A}$ in the Newtonian gauge as,
\begin{equation} \label{eq:11}
    \mathcal{E}=2\Phi, \hspace{2em} \mathcal{A}=-2\Psi
\end{equation}
With this, considering only the scalar modes, the components of the perturbed metric and the tensor field $B_{\mu\nu}$ can be written as,
\begin{equation} \label{eq:12}
    g_{00}=-1-2\Phi, \hspace{2em} g_{0i}=0, \hspace{2em} g_{ij}=a^2\delta_{ij}(1-2\Psi)
\end{equation}
 \begin{equation} \label{eq:13}
    B_{00}=0, \hspace{2em} B_{0i} = a\phi-\partial_iu, \hspace{2em} B_{ij} = \epsilon_{ijk}\partial^kv
\end{equation}
Since, we are at the beginning stage for analyzing scalar perturbations, we will be considering the simplest case for our action that can give viable inflationary solutions, which is the minimally coupled scenario. The minimally coupled action can be written as,
\begin{equation}\label{eq:14} 
     S=\int d^{4}x\sqrt{-g}\left[\frac{R}{2\kappa}-\frac{1}{12}H_{\lambda\mu\nu}H^{\lambda\mu\nu}+\frac{\tau}{2}(\nabla_{\lambda}B^{\lambda\nu})(\nabla_{\mu}{B^{\mu}}_{\nu})-\frac{m^2}{4}B_{\mu\nu}B^{\mu\nu}-\frac{m^2\phi_0^2}{4}\right]
\end{equation}
We substitute the explicit forms of the metric and the tensor field in this minimally coupled action, and perturb the action up to second order in the perturbation variables. We are particularly interested in the second order part of the action since its variation will give rise to the equations of motion which are of first order in the perturbed modes. Initially, we have the four scalar modes $\Phi, \Psi, u$ and $v$. We are making a transformation $u\rightarrow \tilde{u}=\frac{ik}{a\phi}u$. This is to obtain a similar equation structure for the $u$ mode in comparison to the other modes while solving the evolution equations. This will make it easier to find the corresponding solutions and analyze them. Apart from this, we transform our action into its Fourier counterpart with which we will do our analysis.
\subsection{Stability Analysis}
In our previous study on tensor modes, we had looked for the presence of potential pathological instabilities \cite{Aashish:2021gdf}, that could plague our model, but fortunately we could bypass them \cite{Aashish:2019zsy}. We had also performed a stability analysis on the standalone scalar and vector modes coming from the $B_{\mu\nu}$ field and found them to be stable. But it had been mentioned that we could expect pathological instabilities in a more general setup considering the modes from the metric part as well. Keeping this in  mind, we can look at the stability of the scalar modes. Our objective is to check whether ghost and gradient instabilities appear. Ghost instabilities occur when the coefficients of the kinetic terms of the perturbed modes acquire negative coefficients. They make the theories ill defined and cause the energy to be unbounded from below. From the second order action in Fourier space, the relevant terms for finding ghosts can be cast in the form,
\begin{equation}\label{eq:15}
    S_2^{\text{Kin}}=\int \dot{\Delta}^{\dagger}K\dot{\Delta}
\end{equation}
with the vector $\Delta$ consisting of the scalar modes and is given by,
\begin{equation}\label{eq:16}
\renewcommand\arraystretch{1}
    \Delta=\begin{pmatrix}
        \Phi \\
        \Psi \\
        \tilde{u} \\
        v
        
    \end{pmatrix}, \hspace{3em} \Delta^{\dagger}=\begin{pmatrix}
     \Phi^{\dagger} &
        \Psi^{\dagger} &
        \tilde{u}^{\dagger} &
        v^{\dagger}   
    \end{pmatrix}
\end{equation}
$K$ is the square matrix that encapsulates the coefficients of the kinetic terms in our action,
\begin{equation}\label{eq:17}
\renewcommand\arraystretch{1.25}
    K = \begin{bmatrix}
        \frac{3}{2}\tau a^3\phi^2 & \frac{3}{2}\tau a^3\phi^2 & -\frac{1}{2}\tau a^3\phi^2 & 0 \\
        \frac{3}{2}\tau a^3\phi^2 & \frac{3}{2}\tau a^3\phi^2-\frac{3a^3}{\kappa} & -\frac{1}{2}\tau a^3\phi^2 & 0 \\
        -\frac{1}{2}\tau a^3\phi^2 & -\frac{1}{2}\tau a^3\phi^2 & \frac{1}{2}\tau a^3\phi^2 & 0 \\
        0 & 0 & 0 & k^2/2a
    \end{bmatrix}
\end{equation}
We can see that there are no non-dynamical modes in the system. The eigen values of this matrix $K$ can tell us about the ghosts. Approximating the field $\phi$ at its de Sitter value, we can find the  eigen values to be  $-2.588 \ a^3/\kappa$, $0.6566 \ a^3/\kappa$, $0.0980 \ a^3/\kappa$ and $k^2/2a$. Here, we have used a value of $\kappa\tau\phi_0^2=1$, consistent with the constraint on the same from the minimal model. Since one of the eigen values is negative, we encounter a ghost instability with one of the modes. Note that with $\tau =0$, the modes $u$ and $\Phi$, become non-dynamical. We also look for the presence of gradient instability in our model. It can be checked by evaluating the sound speed, $c_s$ for our scalar modes. An imaginary value for the sound speed indicates the gradient instability. To calculate $c_s$ it is necessary to derive the equations of motion for our system. Variation of the perturbed second-order action with respect to each of the modes $\Phi, \Psi$,  $\tilde{u}$ and $v$ will give their corresponding equations of motion. Upon determining them, it is seen that the differential equations for the three modes $\Phi, \Psi$ and $\tilde{u}$ are coupled with each other, while the $v$ mode evolves independently. The equation of motion for $v$ is given as,
\begin{equation}\label{eq:18}
    \ddot{v}-H\dot{v}+\left(\frac{k^2}{a^2}+m^2 \right)v = 0
\end{equation}
whereas the coupled equations for the other modes can be cast into a matrix equation of the form,
\begin{equation}\label{eq:19}
\renewcommand\arraystretch{1}
    \alpha\ddot{\Delta}+\beta\dot{\Delta}+\gamma\Delta = 0 ,\hspace{2em} \text{with} \hspace{2em} \Delta=\begin{pmatrix}
        \Phi \\
        \Psi \\
        \tilde{u} 
        
    \end{pmatrix}
\end{equation}
with the coefficient matrices $\alpha$ and $\beta$ taking the form,
\begin{equation}
\renewcommand\arraystretch{1}
    \alpha= \begin{bmatrix}
        3 & 3  & -1 \\
        3 & 3-6Y^2 & -1  \\
        -1 & -1  & 1 
        
    \end{bmatrix},  \hspace{3em} \beta = H \begin{bmatrix}
        9 & -3+6Y^2  & -1 \\
        21-6Y^2 & 9-18Y^2 & -5  \\
        -5 & -1  & 3 
        
    \end{bmatrix}
\end{equation}
while the components of  $\gamma$ are given as,
\begin{equation}
    \begin{aligned}
          \gamma_{11} &= \frac{k^2}{a^2} + 9H^2Y^2 -\frac{9m^2}{2\tau}-\frac{m^2\theta Y^2}{4\tau} \\
          \gamma_{12} &=  \frac{k^2}{a^2}(1+2Y^2)+ 9H^2Y^2 - \frac{3m^2}{2\tau}-\frac{3m^2\theta Y^2}{4\tau}\\
          \gamma_{13} &=  -\frac{k^2}{a^2}+\frac{m^2}{\tau} \\
          \gamma_{21} &=  \frac{k^2}{a^2}(1+2Y^2)+ 9H^2(4-Y^2) - \frac{3m^2}{2\tau}-\frac{3m^2\theta Y^2}{4\tau}\\
          \gamma_{22} &= \frac{k^2}{a^2}(1-2Y^2)-3H^2(4+3Y^2) + \frac{3m^2}{2\tau}+\frac{3m^2\theta Y^2}{4\tau} \\
          \gamma_{23} &= -\frac{k^2}{a^2}-8H^2+\frac{m^2}{\tau} \\
          \gamma_{31} &= -\frac{k^2}{a^2}-6H^2+\frac{m^2}{\tau} \\
          \gamma_{32} &= -\frac{k^2}{a^2}-6H^2+\frac{m^2}{\tau} \\
          \gamma_{33} &= \frac{k^2}{a^2}+2H^2-\frac{m^2}{\tau}
    \end{aligned}
\end{equation}
Here, $Y^2$ and $\theta$ are background quantities that correspond to $1/\kappa\tau\phi^2$ and $\kappa\tau\phi_0^2$ respectively. A reasonable ansatz for the solutions to eqs. (\ref{eq:18}) and (\ref{eq:19}) in the deep subhorizon limit is to take the modes proportional to $\exp[-i\int^tc_sk/a(t')dt']$. We apply this assumption individually in eqs. (\ref{eq:18}) and (\ref{eq:19}). For the $v$ mode, we have,
\begin{equation}\label{eq:21}
    c_s^2+2i\left(\frac{aH}{k} \right)c_s-\left(1+\frac{m^2a^2}{k^2} \right) = c_s^2+2i\left(\frac{aH}{k} \right)c_s-1-4\tau\left(\frac{ aH}{k} \right)^2 = 0
\end{equation}
To get around gradient instability, one must ensure that $c_s^2$ is positive. Our current region of interest lies in the deep subhorizon, where $k>>aH$. Basically, we are dealing with the high momentum limit. It is possible for instabilities to show up in the low momentum limit, but they have been shown to be recast into Jeans-like instabilities and may be controlled \cite{Gumrukcuoglu:2016jbh}. 
In the limit $k>>aH$, we have,
\begin{equation}\label{eq:22}
    c_s^2=1
\end{equation}
 For the other coupled modes, substituting the ansatz will give rise to an eigen value problem. On evaluation, in the deep subhorizon limit, we obtain,
\begin{equation} \label{eq:23}
    c_s^2 = 1, \frac{1}{2}\left(1\pm\sqrt{1-\frac{4}{3}Y^2}\right)
\end{equation}
For the minimal model, we have $1<4Y^2/3$ and hence the quantity inside the square root becomes negative and $c_s^2$ will turn out to be a complex number with an imaginary part. If we are obtaining $c_s$ as a complex number it shows that the resulting solutions in the deep subhorizon limit may attain exponential growth leading to gradient instability. It is to be noted that with the inclusion of the non-minimal coupling $RB_{\mu\nu}B^{\mu\nu}$, the kinetic matrix $K$ get modified as,
\begin{equation}
    K^{\rm NM} = K^{\rm M} + \frac{\xi}{\kappa\tau}\begin{bmatrix}
        0 & 18\tau a^3\phi^2 & 0 & 0 \\
        18\tau a^3\phi^2 & -18\tau a^3\phi^2 & -6\tau a^3\phi^2 & 0 \\
        0 & -6\tau a^3\phi^2 & 0 & 0 \\
        0 & 0 & 0 & 0
    \end{bmatrix}
\end{equation}
Here $K^{\rm NM}$ and $K^{\rm M}$ represent the kinetic matrices of the non-minimal and minimal models respectively. Approximating de Sitter value with sample values of  $\theta\approx 0.001$ and the coupling strength $\xi/\kappa\tau\approx800$ consistent with our earlier results \cite{Aashish:2021gdf}, the eigen values are now $-1.728\times 10^7 \ a^3/\kappa$, $6.9152\times 10^6 \ a^3/\kappa$ , $216.03 \ a^3/\kappa$ and $k^2/2a$ respectively. Further, 
the matrices $\alpha$, $\beta$ and $\gamma$ get modified as,
\begin{equation}
    \alpha^{\rm NM}=\alpha^{\rm M}+ \frac{\xi}{\kappa\tau}\begin{bmatrix}
        0 & 36  & 0 \\
        36 & -36 & -12  \\
        0 & -12  & 0 
        
    \end{bmatrix},  \hspace{2em} \beta^{\rm NM} = \beta^{\rm M}+ \frac{\xi H}{\kappa\tau} \begin{bmatrix}
        0 & 144  & 12 \\
        72 & -108 & -24  \\
        -12 & -48  & 0 
        
    \end{bmatrix}
\end{equation}
\begin{equation}
\renewcommand\arraystretch{1.5}
    \gamma^{\rm NM}=\gamma^{\rm M}+ \frac{\xi}{\kappa\tau}\begin{bmatrix}
        270H^2-\frac{24k^2}{a^2} & 54H^2+\frac{24k^2}{a^2}  & \frac{4k^2}{a^2}-36H^2 \\
        \frac{24k^2}{a^2}-54H^2 & 18H^2-\frac{36k^2}{a^2} & 12H^2-\frac{8k^2}{a^2}  \\
        \frac{4k^2}{a^2}-72H^2 & -24H^2-\frac{8k^2}{a^2}  & 24H^2
        \end{bmatrix}
\end{equation}
where the superscript M over the matrices represent their minimal model counterparts. Estimating $c_s^2$ for the aforementioned sample values, we obtain,
\begin{equation}
    c_s^2=1, \ 0.1668\pm18.8417i
\end{equation}
We can see that even with the inclusion of the non-minimal coupling term ($RB_{\mu\nu}B^{\mu\nu}$), there is no change in the nature of the terms that determine the stability of the system. Hence, we are not including the non-minimal coupling term for our further analysis.
\subsection{Evolution of scalar modes}
In the previous section, we looked at the potential instabilities that appear among the scalar modes. Now we can find the solutions to their equations of motion to probe the evolutionary characteristics at different stages of inflation. We have already found the equations of interest while dealing with the gradient instabilities. It is seen that the $v$ mode equation is independent while the other modes are entangled with each other. Hence, it is straightforward to first look at the evolution of the $v$ mode. It is advantageous to shift to the conformal time coordinate $\eta$. The derivative with respect to $\eta$ will be represented by a prime over the quantity. Further, owing to the complexity involved in our model, we are interested in the approximation of a simpler de Sitter spacetime, rather than a quasi de Sitter one. This approximation will make the equations look simpler, so that we can get a rough picture of the scenario we are dealing with. 
The equation of motion for the $v$ mode in conformal time looks like,
\begin{equation}\label{eq:24}
    v''-2aHv'+(k^2+m^2a^2)v=0
\end{equation}
We redefine $v$ as $v\rightarrow \tilde{v}=a^{-1}v$. In terms of this new variable we can write,
\begin{equation}\label{eq:25}
    \tilde{v}''+\left[k^2-a^2H^2\left(1-\frac{m^2}{H^2}\right)\right]\tilde{v} = 0
\end{equation}
In the de-Sitter limit we can make the approximation, $aH\approx-1/\eta$. Also using the de Sitter value for $H$ inside the parentheses, we write,
\begin{equation}\label{eq:26}
    \tilde{v}''+\left[k^2-\frac{\omega^2}{\eta^2} \right]\tilde{v}=0, \hspace{2em}{\rm with}\hspace{2em} \omega^2=1-4\tau
\end{equation}
Eq. (\ref{eq:26}) closely resembles that of a harmonic oscillator. So, we can quantize the oscillator by writing $\tilde{v}$ in terms of the creation and annihilation operators $\hat{a}_{\vec{k}}$ and $\hat{a}_{\vec{k}}^\dagger$ respectively as, 
\begin{equation}\label{eq:27}
    \tilde{v}(k,\eta)=\mu(k,\eta)\hat{a}_{\vec{k}}+\mu^*(k,\eta)\hat{a}_{\vec{k}}^{\dagger} 
\end{equation}
The * over the quantity represents the corresponding complex conjugate. Further, the coefficients $\mu(k,\eta)$ satisfy the equation,
\begin{equation}\label{eq:41}
    \mu''+\left[k^2-\frac{\omega^2}{\eta^2} \right]\mu = 0
\end{equation}
For finding these coefficients we can use the Bunch-Davies vacuum state, i.e the state defined as $\hat{a}_{\Vec{k}}|0\rangle=0$ \cite{Kundu:2011sg}. We look at the solutions separately on subhorizon and superhorizon scales. For a particular $k$ mode, the sub horizon limit is characterized by the condition $k>>aH$. Thus, for the subhorizon scales, we have $k^2>>1/\eta^2$. Hence, we can neglect the second term inside the square bracket in eq. (\ref{eq:41}). The solution we obtain is  oscillatory, which, after normalization looks like,
\begin{equation} \label{eq:29}
    \mu(k,\eta)=\frac{1}{\sqrt{2k}} e^{-ik\eta}
\end{equation}
Similarly, for the superhorizon scales, we have $k<<1/|\eta|$. Hence, the second term inside the square brackets dominates and the solution looks like,
\begin{equation} \label{eq:30}
    \mu \propto a^{-(1\pm \sqrt{1+4\omega^2})/2} \hspace{1em} \Longrightarrow \hspace{1em} \mu \propto a^{-(1\pm \sqrt{5-16\tau})/2}
\end{equation}
Thus we have our $v$ mode as,
\begin{equation}\label{eq:31}
    v\propto a^{(1\pm\sqrt{5-16\tau})/2} 
\end{equation}
The nature of the solutions depend on the parameter $\tau$. If $\tau$ were to take a value of $1/4$, then it is possible for $v$ to attain a constant solution on super horizon scales. Now, we try to find the exact solution for the mode. We can define a new variable $p=-k\eta$ and redefine $\mu\rightarrow\tilde{\mu}=p^{-\frac{1}{2}}\mu$. Note that since $\eta$ varies from $-\infty$ to 0 during inflation, the variable $p$ remains positive during the same. Employing these transformations eq. (\ref{eq:41}) can be rewritten into a Bessel differential equation with order parameter $\nu$ as follows,
\begin{equation}\label{eq:32}
    p^2\frac{d^2\tilde{\mu}}{dp^2}+p\frac{d\tilde{\mu}}{dp}+(p^2-\nu^2)\tilde{\mu}=0
\end{equation}
Here $\nu^2=\omega^2+\frac{1}{4}= \frac{1}{4}(5-16\tau)$. The solution of Bessel's differential equation can be written as a sum of the Hankel functions of first and second kind. In our case, we obtain,
\begin{equation}\label{eq:33}
    \mu(p)=\sqrt{p}(A_1H_{\nu}^{(1)}(p)+A_2H_{\nu}^{(2)}(p))
\end{equation}
Matching the subhorizon solutions with the asymptotic forms of the Hankel functions, we find that $A_2$ is 0 (see for reference \cite{Aashish:2021gdf}).
Thus the exact solution in the subhorizon limit will be,
\begin{equation}\label{eq:34} 
    \mu(k,\eta)=\frac{\sqrt{\pi}}{2}e^{i\left(\nu+\frac{1}{2}\right)\frac{\pi}{2}}\sqrt{-\eta} H_{\nu}^{(1)}(-k\eta)
\end{equation}
Hence, the exact solution for $v$ mode in the superhorizon limit will look like,
\begin{equation}\label{eq:35}
    v(k,\eta) = \frac{\Gamma(\nu)}{\Gamma(3/2)}e^{i\left(\nu-\frac{1}{2}\right)\frac{\pi}{2}}2^{\nu-2}\sqrt{k} \ (-k\eta)^{-\frac{1}{2}-\nu} \hspace{2em} \nu=\pm\frac{\sqrt{5-16\tau}}{2}
\end{equation}
where $\Gamma$ is the Euler function. Note that even though a $\tau$ value of 1/4 can give a constant solution in proper time for the $v$ mode on superhorizon scales, the r.m.s amplitude $k^{\frac{3}{2}}v$ cannot be made independent of scales. This is primarily due to the appearance of a negative coefficient for the $\dot{v}$ term in eq. (\ref{eq:18}). Due to this negative term, the relation between $v$ and $\tilde{v}$ flips while comparing the similar scenario with the tensor mode $h_e$. Hence, we are getting a $\sqrt{k}$ in place of the $k^{-3/2}$ in the term apart from $-k\eta$ for the superhorizon solution. The addition of non-minimal couplings with Ricci scalar or tensor that we studied cannot make the power spectrum scale invariant due to the lack of $B_{\mu\nu}$ derivatives in the coupling. Only a gravity coupling that contains the $B_{\mu\nu}$ derivatives can modify the $\dot{v}$ term. But such couplings will introduce higher derivatives in the theory making it difficult to consider their inclusion. 
However, using suitable combinations that could involve clever cancellation of the higher derivatives may be used. But such inclusions will modify all the background and perturbative dynamics we have considered so far and hence will be out of the scope of this work. However, the effect of such modifications in the context of antisymmetric tensor field inflation is worth exploring, and we leave it for future studies. 
Now, we can look at the evolutionary behavior of the other modes. Since they are coupled, we need to analyse them together. It is convenient to transform our equations in terms of the variable $p=-k\eta$ which makes it easier to distinguish the behaviour at different regimes. The coupled equations now look like, 
\begin{equation}
    \begin{aligned}
      p^2(3(\ddot{\Phi}+\ddot{\Psi})-\ddot{\tilde{u}}) -6p(\dot{\Phi}+\dot{\Psi}(Y^2-1)) + 
      \tilde{u}(4-p^2)&+ \\ \Psi(p^2(1+2Y^2)-3Y^2(\theta-3)-6)+\Phi(p^2-Y^2(\theta-9)-18) &= 0
    \end{aligned}
\end{equation}
\begin{equation}
    \begin{aligned}
    p^2(3\ddot{\Phi}+3\ddot{\Psi}(1-2Y^2)-\ddot{\tilde{u}})+2p(2\dot{\tilde{u}}+(6Y^2-3)\dot{\Psi}+(3Y^2-9)\dot{\Phi})-\tilde{u}(4+x^2) &+ \\
    \Psi(p^2(1-2Y^2)+3Y^2(\theta-3)-6) + 
    \Phi(30-3Y^2(3+\theta)+p^2(1+2Y^2)) &= 0
    \end{aligned}
\end{equation}
\begin{equation}\label{eq:38}
    p^2(\ddot{\tilde{u}}-\ddot{\Psi}+\ddot{\Phi}) + 2p(2\dot{\Phi}-\dot{\tilde{u}})+\tilde{u}(p^2-2)+\Psi(2-p^2)-\Phi(2+p^2) = 0
\end{equation}
Note that here the dot over the quantity indicates its corresponding derivative with respect to the variable $p$. We are interested in the behaviour of these modes at different stages in its evolution. For a mode with a particular $k$ value the condition $p>>1$ marks the subhorizon limit where as the opposite $p<<1$ is true for the super horizon case. A $p$ value that is closer to 1 represents the horizon crossing of the mode. The evolution of these equations begin from the deep subhorizon limit, where these modes are very small compared to the horizon scale. In this limit, we can approximate our system of equations as,
 \begin{equation}\label{eq4.49}
    3\ddot{\Psi}+3\ddot{\Phi}-\ddot{\tilde{u}}+\Psi(2Y^2+1)+\Phi-\tilde{u}=0    
\end{equation}
\begin{equation}\label{eq4.50}
    3\ddot{\Phi}+3\ddot{\Psi}(1-2Y^2)-\ddot{\tilde{u}}+\Psi(1-2Y^2)+\Phi(1+2Y^2)-\tilde{u}=0
\end{equation}
\begin{equation}\label{eq4.51}
    \ddot{\Psi}+\ddot{\Phi}-\ddot{\tilde{u}}+\Psi+\Phi-\tilde{u}=0
\end{equation}
Subtracting eq. (\ref{eq4.51}) from eqs. (\ref{eq4.49}) and (\ref{eq4.50})  reduces the dimensionality of the system, with the resulting  two equations consisting entirely of the modes $\Phi$ and $\Psi$. 
\begin{equation}\label{eq:42}
    \ddot{\Psi}+\ddot{\Phi}+Y^2\Psi=0
\end{equation}
\begin{equation}\label{eq:43}
    \ddot{\Phi}+\ddot{\Psi}(1-3Y^2)+Y^2(\Phi-\Psi)=0
\end{equation}
We can solve these equations to find the modes $\Phi$ and $\Psi$ and use them to find $\tilde{u}$. After performing these operations, the corresponding analytical solutions we obtain for the coupled modes $\Phi$, $\Psi$ and $\tilde{u}$ look like, 
\begin{multline}\label{eq:44}
   \Phi = e^{-0.572p} \bigg[(0.5 c_1-0.185 c_2-1.26 c_4) \cos (0.91p)+0.5 (c_1+0.369 c_2) e^{1.14p} \cos (0.91p) + \\ 0.433 c_2 \sin (0.91p) - 1.6 c_3 e^{1.14p} \sin (0.91p)+ 1.6 c_3 \sin (0.91p)-0.794 c_4 e^{1.14p} \sin (0.91p)      \\ +2.52 c_4  e^{1.14p} \cos (0.91p)+  0.08 c_1 e^{1.14p} \sin (0.91p)+0.433 c_2 e^{1.14p} \sin
   (0.91p) - \\   0.08 c_1 \sin (0.91p) -0.794 c_4 \sin (0.91p)\bigg]
\end{multline}
\begin{multline}\label{eq:45}
    \Psi=e^{-0.572p} \bigg[-0.126 (c_2-3.97 c_3-2.47 c_4) e^{1.14p} \cos (0.91p)+(0.126 c_2+0.5 c_3) \cos (0.91p) \\ -0.311 c_4 \cos (0.91p) + 0.16 c_1 e^{1.14p} \sin (0.91p)-0.16 c_1 \sin (0.91p)+0.0793 c_2 e^{1.14p} \sin
   (0.91p) \\ +0.0793 c_2 \sin (0.91p) -0.08 c_3 e^{1.14p} \sin (0.91p) +0.08 c_3 \sin (0.91p)  +0.354 c_4 e^{1.14p} \sin (0.91p)\\ +0.354 c_4 \sin (0.91p)\bigg]
\end{multline}
\begin{multline}\label{eq:46}
    \tilde{u}=e^{-0.572p} \bigg[(0.5 c_1-0.059 c_2+0.5 c_3-1.57 c_4) \cos (0.91p)+0.5 (c_1+0.117 c_2) e^{1.14p} \cos (0.91p)\\ + (c_3+3.14 c_4)\cos (0.91p) + c_5 e^{0.572p} \cos (0.91p)+0.24 c_1 e^{1.14p} \sin (0.91p)-0.24
   c_1 \sin (0.91p)+ \\ 0.513 c_2 e^{1.14p} \sin (0.91p)+0.513 c_2 \sin (0.91p)  -1.68 c_3 e^{1.14p} \sin (0.91p)+1.68 c_3 \sin (0.91p)\\ -0.44 c_4 e^{1.14p} \sin (0.91p)-0.44 c_4 \sin (0.91
  p)+ c_6 e^{0.572p} \sin (0.91p)\bigg]
\end{multline}
Upon inspecting them, we can see that the solutions are oscillatory and consists of rapidly growing or decaying modes. We have already expected this behavior from the analysis on gradient instabilities. As the modes grow larger rapidly they no longer retain their nature as perturbations. Hence, the equations we used to govern their dynamics will no longer be applicable, rendering the modes out of control. In order to prevent the growth of such modes, which can potentially break down the perturbation theory, we can set the coefficients of the exponential terms with positive arguments to be 0. This leads to the following constraints on the integration constants $c_i$ given in eqs. (\ref{eq60}), (\ref{eq61}) and (\ref{eq62}). If we can impose these constraints on the boundary values $c_i$ in the subhorizon solutions, we can suppress the rapidly growing modes, thereby finding a solution to the gradient instability.
\begin{equation}\label{eq60}
    (0.5 c_1+0.185 c_2+1.26 c_4) +(0.08 c_1+0.433 c_2-1.6 c_3-0.794 c_4) \tan (0.91p) = 0
\end{equation}
\begin{equation}\label{eq61}
     (-0.126 c_2+0.5 c_3+0.311 c_4) +(0.16 c_1+0.0793 c_2-0.08 c_3+0.354 c_4) \tan (0.91p) = 0
\end{equation}
\begin{equation}\label{eq62}
     (0.5 c_1+0.059 c_2+0.5 c_3+1.57 c_4) +(0.24 c_1+0.513 c_2-1.68 c_3-0.44 c_4) \tan (0.91p) = 0
\end{equation}
\section{Vector modes}\label{sec:4}
Now we try to look at the vector modes in our model. Unlike the standard scalar field model where the vector modes originate only from the metric and decay off during expansion, here there are contributions from our external field as well. Hence, they can be dynamically sourced, which may allow them to survive and grow. The presence of such modes can source vorticity during inflation, which is a proposed mechanism for the generation of primordial magnetic fields \cite{Naoz:2013wla,Demozzi:2009fu,Basak:2014qea}. We have four divergence free vector fields $\mathcal{C}_i, \ \mathcal{G}_i, \ U_i$ and $V_i$ in our model. Since, there are two independent components in $\xi_i^V$, there are two gauge degrees of freedom in the vector sector. Hence, we can choose $\xi_i^V$ such that either $\mathcal{C}_i$ or $\mathcal{G}_i$ vanishes since the other two are gauge invariant for our choice. For the gauge choice with vanishing $\mathcal{C}_i$, the $\mathcal{G}_i$ modes are seen to not harbor any kinetic terms, making them non-dynamical. This may induce instabilities in the propagation of the other modes. Thus it is better to resort to the usage of the gauge with $\mathcal{G}_i=0$. With this condition, we substitute the explicit forms of the metric and tensor field in our minimally coupled action given in eq. (\ref{eq:14}). We can calculate the equations of motion from the second order part of the perturbed action. Analogous to what we did for the scalar modes, we first look at the stability arguments and then move on to the dynamics of the fields.
\subsection{Stability Analysis}
First, we will look for the ghost instabilities in the model. The Kinetic matrix $K$ in eq. (\ref{eq:15}) can be written for the vector modes as,
\begin{equation}
\renewcommand\arraystretch{1.25}
    K = \frac{k^2a^3}{\kappa}\begin{bmatrix}
        \frac{1}{4}+\phi^2_{\kappa\tau} & \phi^2_{\kappa\tau}/2 & -i\phi^2_{\kappa\tau}/2ka\phi & 0 & 0 & 0 \\
        \phi^2_{\kappa\tau}/2 &  \frac{1}{4} +\phi^2_{\kappa\tau} & 0 & -i\phi^2_{\kappa\tau}/ka\phi & 0 & 0 \\
        i\phi^2_{\kappa\tau}/2ka\phi & 0 & \kappa\tau/2k^2a^2 & 0 & 0 & 0\\
        0 & i\phi^2_{\kappa\tau}/2ka\phi & 0 & \kappa\tau/2k^2a^2 & 0 & 0 \\
        0 & 0 & 0 & 0 & \kappa/2k^2a^4 & 0 \\
        0 & 0 & 0 & 0 & 0 & \kappa/2k^2a^4
    \end{bmatrix}
\end{equation}
with $\phi^2_{\kappa\tau} \equiv\kappa\tau\phi^2$. The $\Delta$ vector is now  defined as,
\begin{equation}\label{eq:39}
\renewcommand\arraystretch{1}
    \Delta=\begin{bmatrix}
        C_x \\
        C_y \\
        U_x \\
        U_y \\
        V_x \\
        V_y
        
    \end{bmatrix} 
    ,\hspace{4em} 
    \Delta^\dagger = \begin{bmatrix}
        C_x^\dagger &
        C_y^\dagger &
        U_x^\dagger &
        U_y^\dagger &
        V_x^\dagger &
        V_y^\dagger
    \end{bmatrix}
\end{equation}
We can see that all the modes are dynamical for this configuration. Further, the $\tau$ term is a necessary requirement for making the $U_i$ modes dynamical. The eigen values of the kinetic matrix are given as, \\
\begin{equation}
    \frac{\kappa\tau+k^2a^2\pm\sqrt{k^2\tau^2+k^4a^4}}{4\kappa/a}, \hspace{1.5em} \frac{\kappa\tau+2k^2a^2\pm\sqrt{k^2\tau^2-2k^2\kappa\tau a^2+4k^4a^4}}{4\kappa/a}, \hspace{1.5em} \frac{1}{2a}, \hspace{1.5em} \frac{1}{2a}
\end{equation} \\
For $\tau>0$, we always have $\kappa\tau+k^2a^2>\sqrt{k^2\tau^2+k^4a^4}$ and $\kappa\tau+2k^2a^2>\sqrt{k^2\tau^2-2k^2\kappa\tau a^2+4k^4a^4}$. Therefore all the eigen values of $K$ are positive and therefore vector modes do not encounter ghost instabilities. Now we can probe the gradient instability by finding the $c_s$ value from the equations of motion. Upon derivation, all the six vector modes are seen to get entangled with each other in their equations of motion. The coupled differential equations for the vector modes can be written as,
\begin{equation}\label{eq4.63}
    \frac{k^2 \phi}{4 a^2}( C_x+ C_y) +H^2\phi(2C_x+C_y) + \frac{3} {4}H\phi(\dot{C}_y+3\dot{C}_x)  + \frac{3H}{4 k a}i    \dot{U}_x + \frac{\dot{V}_y}{4 a^3}+\frac{\phi}{4}(\ddot{C}_y+3\ddot{C}_x) + \frac{i  \ddot{U}_x}{4 k a} = 0
\end{equation}
\begin{equation}\label{eq4.65}
    2 C_x      H^2  \phi 
+ \left(\frac{k}{\tau a^3}+\frac{4H^2}{ka}\right)iU_x + \frac{2      H    V_y}{a^3} 
-   H  \phi   \dot{C}_x 
- \frac{Hi  \dot{U}_x}{  k    a} -\frac{\dot{V}_y}{\tau a^3}(1+\tau) - \phi\ddot{C}_x - \frac{i\ddot{U}_x}{ka} = 0
\end{equation}
\begin{equation}\label{eq4.67}
    kHiU_y(\tau-1)+\left(aH^2-\frac{k^2}{a^2} \right)\tau V_x + k^2a\tau\phi\dot{C}_y+k(1+\tau)i\dot{U_y} - H  \dot{V}_x 
+ \ddot{V}_x = 0
\end{equation}
where we have used the de Sitter values of the background fields when necessary. The remaining three equations can be obtained by making the replacements $U_x\rightarrow U_y$, $C_x\rightarrow C_y$, $V_x\rightarrow -V_y$ and vice versa, in each the above equations.
Now, taking the ansatz $\exp[-i\int^tc_sk/a(t')dt']$ for the vector modes, we can solve for the $c_s^2$ values in the deep subhorizon limit, which are obtained as,
\begin{equation}\label{eq69}
    c_s^2= \ 0,\ 1, \ 1, \ 1,\ 1,\ \frac{2}{3}
\end{equation}
Since, we are getting real valued solutions for $c_s$, unlike the scalar modes, the vector modes are free from gradient instabilities in the subhorizon limit. Hence, we can now move to the analysis regarding their evolutionary behavior.
\subsection{Evolutionary behavior}
Having performed the generic stability analysis, we now turn our attention to the fate of the vector modes during the course of their evolution. We can transform our equations in terms of the variable $p=-k\eta$. Since the six differential equations that govern the dynamics of the vector modes are coupled with each other, finding an analytical solution to these equations is a nearly impossible task. Hence, we can only attempt to solve these equations numerically. First, we look at the deep subhorizon limit, where the modes start to evolve from. The equations for the vector modes can be rewritten as a system of 12 first order differential equations for the subhorizon limit which are given in Appendix \ref{app_vec} . A $p$ value much greater than 1 denotes the subhorizon limit. In our numerical computations, for approximating the subhorizon limit, we can evolve the modes starting from  $p\approx100$. For the initial condition, we employ the Bunch-Davies vacuum condition, according to which the modes are $\sim e^{ip}/\sqrt{2k}$. Applying this initial condition, we can evolve the real and imaginary parts separately. 
\begin{figure}[h!]
    \centering
    \includegraphics[width=0.9\linewidth,height=0.38\linewidth]{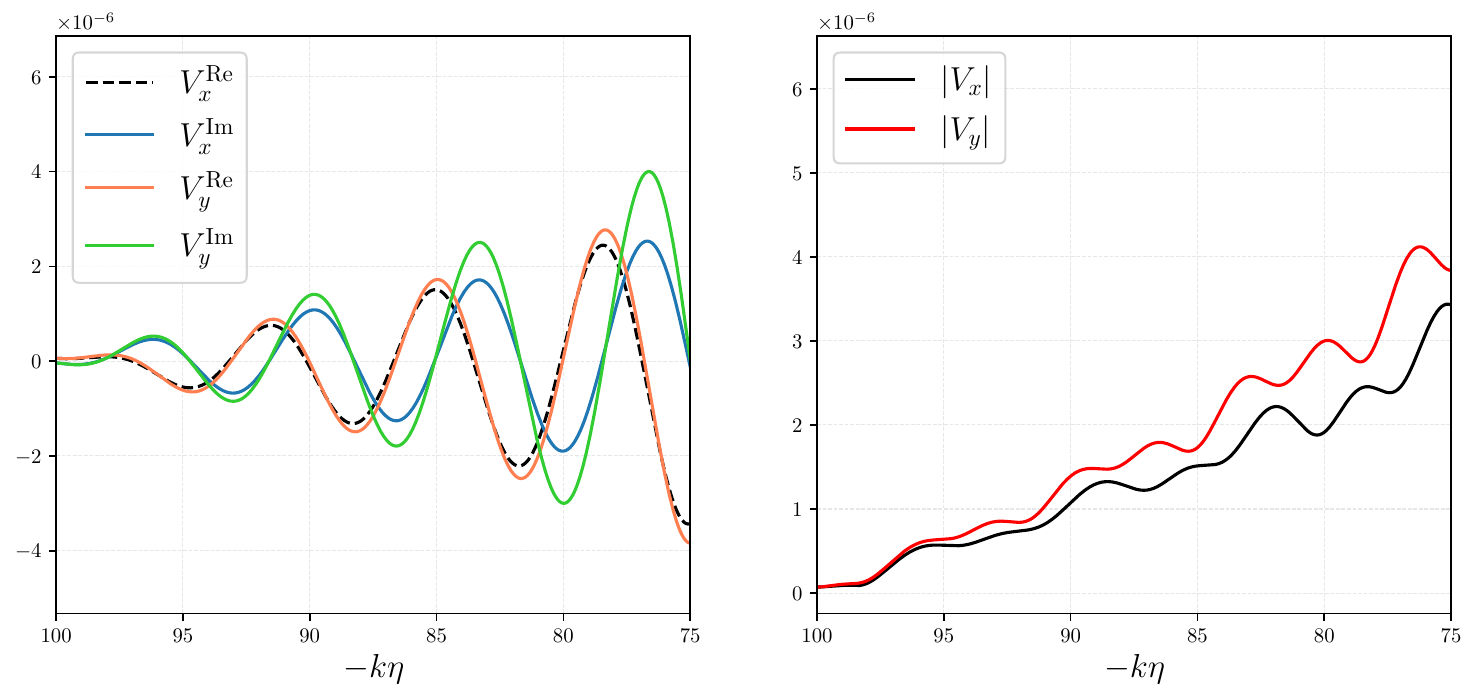}
    \caption{The numerical evolution of the real and imaginary parts of the $V_x$ and $V_y$ modes in the subhorizon limit is shown in the left panel for a $k$ value equal to 10. The right panel gives the modulus value of the corresponding vector modes. We have used a $\tau$ value of the order 1.}
    \label{fig:4.1}
\end{figure}
\begin{figure}[h!]
    \centering
    \includegraphics[width=.9\linewidth,height=0.38\linewidth]{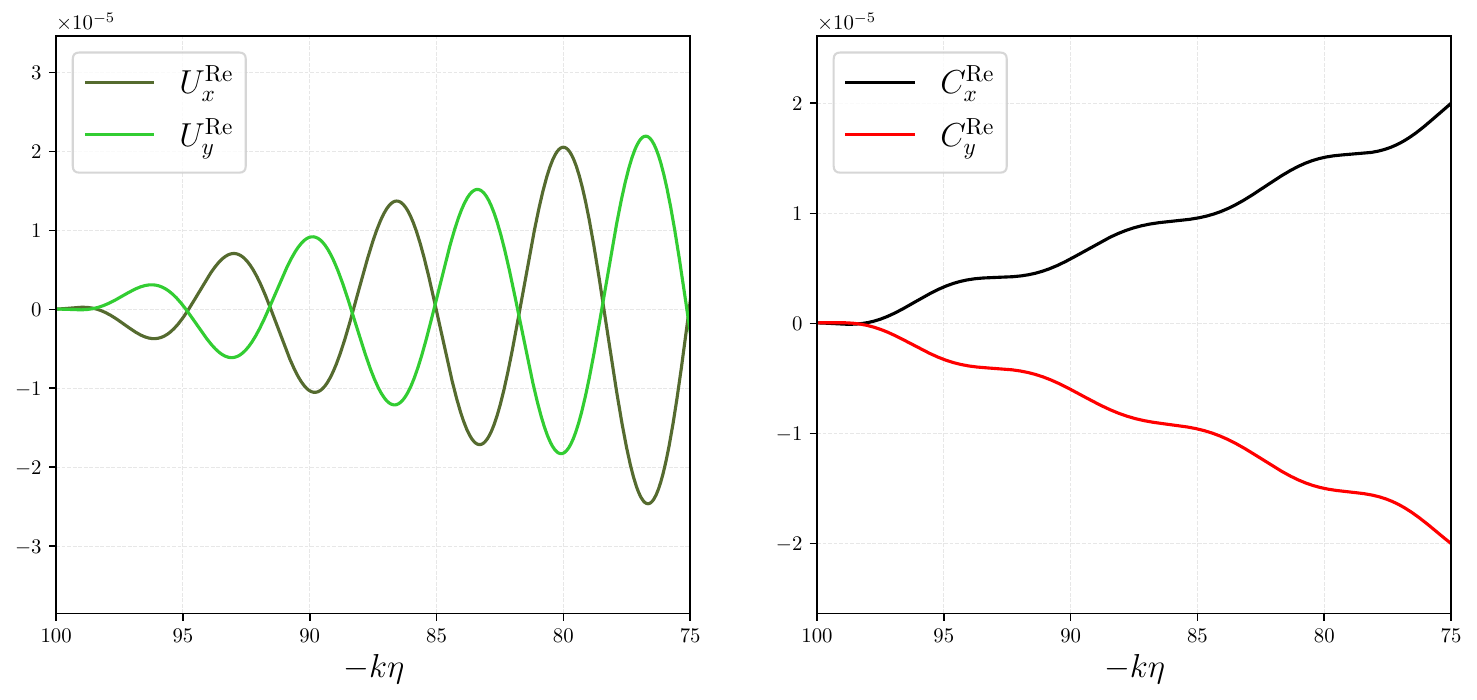}
    \caption{The numerical evolution of the real parts of $U_i$ modes are shown in the left panel, while the right panel showcases the behavior of the corresponding part of the $C_i$ modes. The value of $k$ is taken to be 10, while $\tau$ is of the order 1.}
    \label{fig:4.2}
\end{figure}
 The numerical solutions are plotted in figures (\ref{fig:4.1}) and (\ref{fig:4.2}). For the numerical computation, we have scaled the initial amplitude by a factor of $10^{-7}$. 
We can see that the modes are oscillatory, but grows with time. Even though they are growing, the growth is not rapid to lose their perturbative nature. So, these modes are under control as suggested by our analysis on gradient instabilities in the deep subhorizon limit. But once they pass the subhorizon limit and move towards the horizon, the above calculations will not remain valid. The terms which we neglected in the subhorizon limit can now come into the picture and may cause instabilities that may lead to rapid growth of the modes. Hence, to probe a more complete behavior we will need to solve the full equations and evolve the vector modes from the subhorizon regime all the way towards horizon crossing ($p\sim1$). 
\begin{figure}[h!]
    \centering
    \includegraphics[width=0.6\linewidth,height=0.40\linewidth]{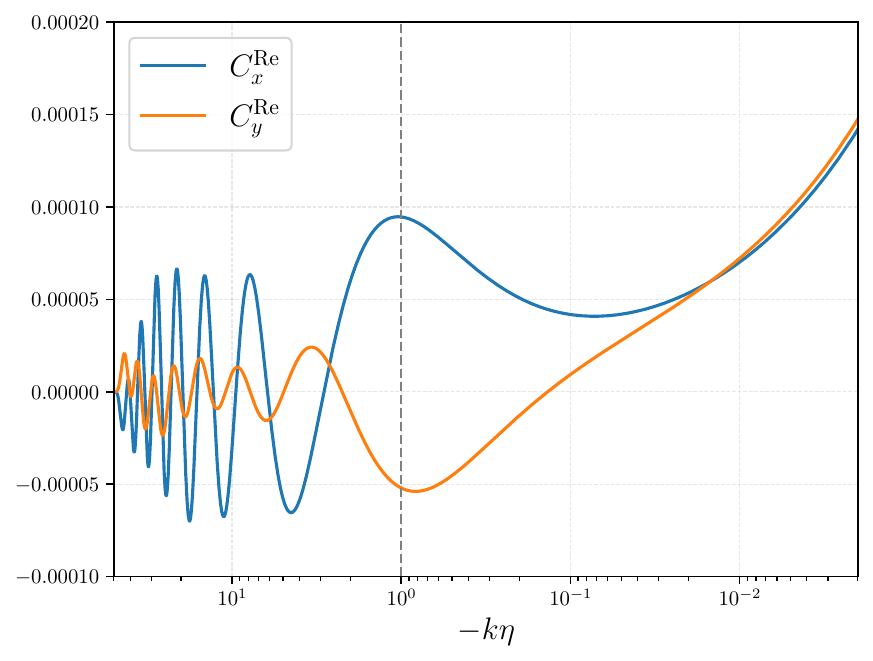}
    \caption{The plot shows the evolution of the real parts of the $C_x$ and $C_y$ modes. The $X$ axis is on logarithmic scale. The dashed vertical line represents the horizon crossing for the modes. $k$, $\tau$ values are 10 and $10^{-7}$ respectively.}
    \label{fig:4.3}
\end{figure} 
\begin{figure}[h!]
    \centering
    \includegraphics[width=.9\linewidth,height=0.38\linewidth]{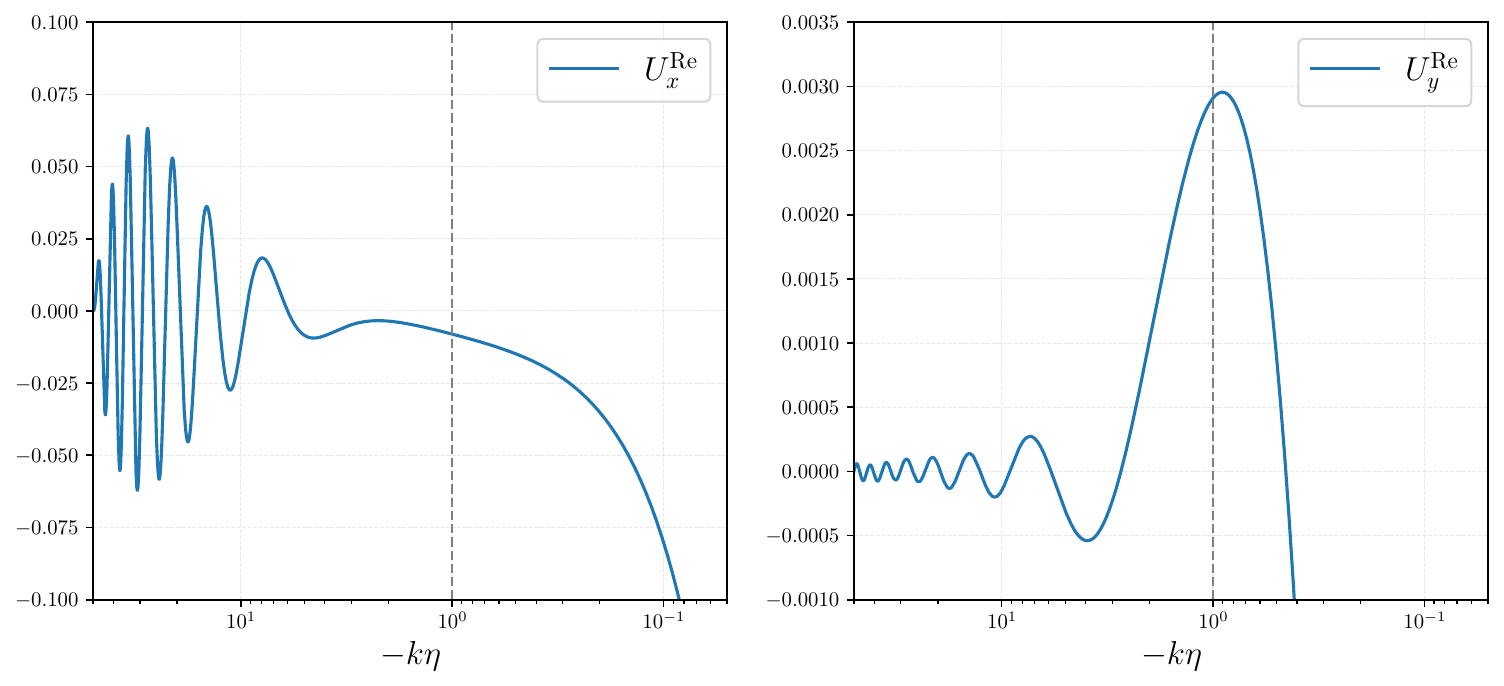}
    \caption{The numerical evolution of the real part of the $U_x$ and $U_y$ modes are shown respectively in the left and right panels. The $X$ axis is on logarithmic scale. The dashed vertical line represents the horizon crossing for the modes. $k$, $\tau$ values are 10 and $10^{-7}$ respectively.}
    \label{fig:4.4}
\end{figure}
\begin{figure}[h!]
    \centering
    \includegraphics[width=.9\linewidth,height=0.38\linewidth]{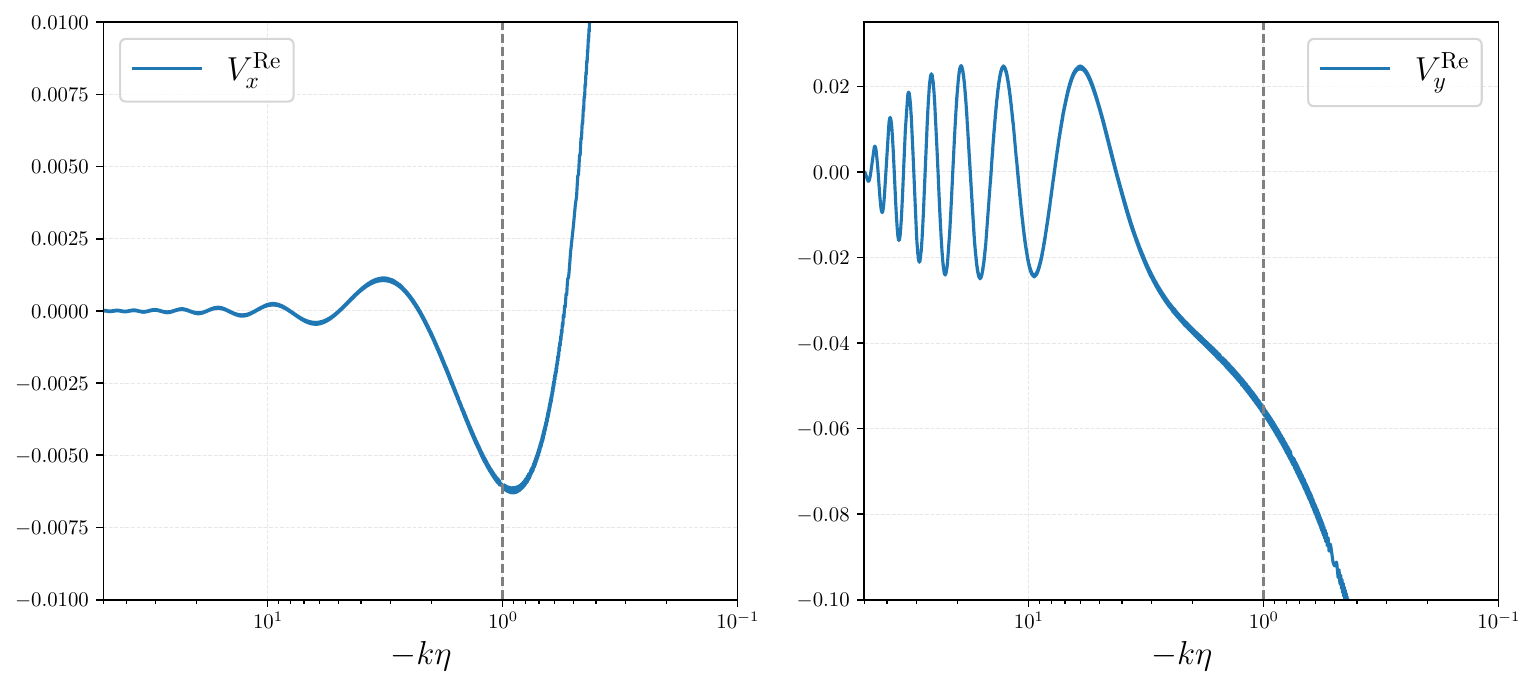}
    \caption{The numerical evolution of the real part of the $V_x$ and $V_y$ modes are shown respectively in the left and right panels. The $X$ axis is on logarithmic scale. The dashed vertical line represents the horizon crossing for the modes. $k$, $\tau$ values are 10 and $10^{-7}$ respectively.}
    \label{fig:4.5}
\end{figure}
The system of the complete first order differential equations without any approximation is also given in Appendix \ref{app_vec}. The evolution of the modes depends on the parameter $\tau$. For a $\tau$ value of the order of 1, even before crossing the horizon, the oscillating vector modes grow out of control such that they are no longer smaller than 1. Hence, the linear perturbation theory breaks down, and the coupled differential equations we derived will no longer be valid. However, by increasing the value of $\tau$, we can keep the vector mode oscillations under control. The evolutionary tracks of the vector modes for a $\tau$ value of the order of $10^{7}$ is shown in figures (\ref{fig:4.3}), (\ref{fig:4.4}) and (\ref{fig:4.5}).
The $X$ axis is displayed on a logarithmic scale to clearly illustrate the transition from subhorizon to superhorizon regime. We can see from the plots that although the modes are under control, after horizon crossing, they are seen to grow. This suggests that the vector modes in our theory can only retain their oscillatory perturbative nature till they cross the horizon. On superhorizon scales, the numerical solutions are exponentially growing rendering instabilities in the perturbations.
\section{Conclusion}\label{sec:5}
As a subsequent step in building the cosmological perturbation theory for our tensor field inflation model, we try to analyze the plausible scalar and vector modes. Since we are at the preliminary stage, we look at the simplest model that can give inflationary solutions, which is the minimal model with a single antisymmetric tensor field, studied in \cite{Aashish:2021gdf}. We can decompose the perturbations into scalar, vector and tensor parts and study them separately at linear order. The perturbations from the antisymmetric tensor field can be decomposed into scalar and vector parts, while the metric perturbations consist of all the three modes. The tensor modes in the context of this model has already been studied in \cite{Aashish:2021gdf}. Upon analyzing the scalar modes, we could see that one of the modes that come from $B_{\mu\nu}$ decouples from the others and evolves separately. This mode on evaluation is seen to give a solution that is constant in proper time on superhorizon scales, but gives a scale dependent power spectrum. The other coupled modes encounter exponential growth or decay in the subhorizon limit. In order to preserve the perturbative nature of the theory, we choose the integration constants in such a way that the growing modes are eliminated from the general solution. Performing the generic stability analysis, the scalar modes are seen to harbor ghost and gradient instabilities in some of the modes, while vector modes lack such instabilities. All the six vector modes are coupled with each other in their equations of motion. Upon numerical evaluation, these modes are seen to oscillate and grow in a controlled rate till horizon crossing, but on superhorizon scales, they encounter exponential growth indicating the presence of instabilities at such scales. An obvious next step in this analysis will be to address the instabilities that occur in this model, particularly in the case of scalar modes. Even with the inclusion of the non-minimal coupling with Ricci scalar studied in \cite{Aashish:2021gdf}, there is no change in the nature of the terms that determine the stability of the system. But they may be alleviated by the introduction of a non-minimal coupling that associates a kinetic term of the field $B_{\mu\nu}$. Further, the tensor field action we considered in our works consisted of parity even terms. It will we interesting to look at the effects of parity-odd terms involving both $B_{\mu\nu}$ and its dual tensor $\mathfrak{B}_{\mu\nu}$.

\bibliographystyle{ieeetr}
\bibliography{refs}

\begin{thebibliography}{10}

\bibitem{Riotto:2002yw}
A.~Riotto, ``{Inflation and the theory of cosmological perturbations},'' {\em ICTP Lect. Notes Ser.}, vol.~14, pp.~317--413, 2003.

\bibitem{Enqvist:2019jkb}
K.~Enqvist, T.~Sawala, and T.~Takahashi, ``{Structure formation with two periods of inflation: beyond PLaIn $\Lambda$CDM},'' {\em JCAP}, vol.~10, p.~053, 2020.

\bibitem{Battye:1998xe}
R.~A. Battye and J.~Weller, ``{Cosmic structure formation in hybrid inflation models},'' {\em Phys. Rev. D}, vol.~61, p.~043501, 2000.

\bibitem{Abedi:2016sks}
H.~Abedi and A.~M. Abbassi, ``{Primordial perturbations in multi-scalar inflation},'' {\em JCAP}, vol.~07, p.~049, 2017.

\bibitem{Kodama:2021yrm}
T.~{Kodama} and T.~{Takahashi}, ``{Relaxing inflation models with non-minimal coupling: A general study},'' {\em arXiv e-prints}, p.~arXiv:2112.05283, Dec. 2021.

\bibitem{Wands:2007bd}
D.~Wands, ``{Multiple field inflation},'' {\em Lect. Notes Phys.}, vol.~738, pp.~275--304, 2008.

\bibitem{Gong:2006zp}
J.-O. Gong, ``{End of multi-field inflation and the perturbation spectrum},'' {\em Phys. Rev. D}, vol.~75, p.~043502, 2007.

\bibitem{Ohashi:2011na}
J.~Ohashi and S.~Tsujikawa, ``{Observational constraints on assisted k-inflation},'' {\em Phys. Rev. D}, vol.~83, p.~103522, 2011.

\bibitem{Vazquez:2018qdg}
J.~A. V'azquez, L.~E. Padilla, and T.~Matos, ``Inflationary cosmology: from theory to observations,'' {\em Revista Mexicana de F{\'i}sica E}, 2020.

\bibitem{Bartolo:2021wpt}
N.~{Bartolo}, A.~{Ganz}, and S.~{Matarrese}, ``{Cuscuton Inflation},'' {\em arXiv e-prints}, p.~arXiv:2111.06794, Nov. 2021.

\bibitem{Starobinsky:1980te}
A.~A. Starobinsky, ``{A New Type of Isotropic Cosmological Models Without Singularity},'' {\em Phys. Lett. B}, vol.~91, pp.~99--102, 1980.

\bibitem{Inagaki:2019hmm}
T.~Inagaki and H.~Sakamoto, ``{Exploring the inflation of $F(R)$ gravity},'' {\em Int. J. Mod. Phys. D}, vol.~29, no.~02, p.~2050012, 2020.

\bibitem{Zhang:2021ppy}
X.~{Zhang}, C.-Y. {Chen}, and Y.~{Reyimuaji}, ``{A new modified gravity framework to rescue inflationary models},'' {\em arXiv e-prints}, p.~arXiv:2108.07546, Aug. 2021.

\bibitem{Sangtawee:2021mhz}
J.~Sangtawee and K.~Karwan, ``{Inflationary model in minimally modified gravity theories},'' {\em Phys. Rev. D}, vol.~104, no.~2, p.~023511, 2021.

\bibitem{Bamba:2015uma}
K.~Bamba and S.~D. Odintsov, ``{Inflationary cosmology in modified gravity theories},'' {\em Symmetry}, vol.~7, no.~1, pp.~220--240, 2015.

\bibitem{Bhattacharjee:2021kar}
S.~Bhattacharjee, ``Inflation in mimetic f(r,t) gravity,'' {\em New Astronomy}, vol.~90, p.~101657, 2022.

\bibitem{Planck:2018vyg}
N.~Aghanim {\em et~al.}, ``{Planck 2018 results. VI. Cosmological parameters},'' {\em Astron. Astrophys.}, vol.~641, p.~A6, 2020.
\newblock [Erratum: Astron.Astrophys. 652, C4 (2021)].

\bibitem{Planck:2019nip}
N.~Aghanim {\em et~al.}, ``{Planck 2018 results. V. CMB power spectra and likelihoods},'' {\em Astron. Astrophys.}, vol.~641, p.~A5, 2020.

\bibitem{Planck:2018lbu}
N.~Aghanim {\em et~al.}, ``{Planck 2018 results. VIII. Gravitational lensing},'' {\em Astron. Astrophys.}, vol.~641, p.~A8, 2020.

\bibitem{SPT-3G:2021eoc}
D.~Dutcher {\em et~al.}, ``{Measurements of the E-mode polarization and temperature-E-mode correlation of the CMB from SPT-3G 2018 data},'' {\em Phys. Rev. D}, vol.~104, no.~2, p.~022003, 2021.

\bibitem{ACT:2020frw}
S.~K. Choi {\em et~al.}, ``{The Atacama Cosmology Telescope: a measurement of the Cosmic Microwave Background power spectra at 98 and 150 GHz},'' {\em JCAP}, vol.~12, p.~045, 2020.

\bibitem{WMAP:2012nax}
G.~Hinshaw {\em et~al.}, ``{Nine-Year Wilkinson Microwave Anisotropy Probe (WMAP) Observations: Cosmological Parameter Results},'' {\em Astrophys. J. Suppl.}, vol.~208, p.~19, 2013.

\bibitem{Planck:2018nkj}
N.~Aghanim {\em et~al.}, ``{Planck 2018 results. I. Overview and the cosmological legacy of Planck},'' {\em Astron. Astrophys.}, vol.~641, p.~A1, 2020.

\bibitem{Lemos:2023rdh}
P.~Lemos and P.~Shah, ``{The Cosmic Microwave Background and $H_0$},'' 7 2023.

\bibitem{Tristram:2007zz}
M.~Tristram and K.~Ganga, ``{Data analysis methods for the cosmic microwave background},'' {\em Rept. Prog. Phys.}, vol.~70, p.~899, 2007.

\bibitem{SPT:2019nip}
J.~T. Sayre {\em et~al.}, ``{Measurements of B-mode Polarization of the Cosmic Microwave Background from 500 Square Degrees of SPTpol Data},'' {\em Phys. Rev. D}, vol.~101, no.~12, p.~122003, 2020.

\bibitem{SPTpol:2013omd}
D.~Hanson {\em et~al.}, ``{Detection of B-mode Polarization in the Cosmic Microwave Background with Data from the South Pole Telescope},'' {\em Phys. Rev. Lett.}, vol.~111, no.~14, p.~141301, 2013.

\bibitem{Bartolo:2019eac}
N.~Bartolo, A.~Hoseinpour, S.~Matarrese, G.~Orlando, and M.~Zarei, ``{CMB Circular and B-mode Polarization from New Interactions},'' {\em Phys. Rev. D}, vol.~100, no.~4, p.~043516, 2019.

\bibitem{DAmico:2022gki}
G.~D'Amico, M.~Lewandowski, L.~Senatore, and P.~Zhang, ``{Limits on primordial non-Gaussianities from BOSS galaxy-clustering data},'' {\em Phys. Rev. D}, vol.~111, no.~6, p.~063514, 2025.

\bibitem{Rezaie:2023lvi}
M.~Rezaie {\em et~al.}, ``{Local primordial non-Gaussianity from the large-scale clustering of photometric DESI luminous red galaxies},'' {\em Mon. Not. Roy. Astron. Soc.}, vol.~532, no.~2, pp.~1902--1928, 2024.

\bibitem{Chaussidon:2024qni}
E.~Chaussidon {\em et~al.}, ``{Constraining primordial non-Gaussianity with DESI 2024 LRG and QSO samples},'' 11 2024.

\bibitem{Iacconi:2019vgc}
L.~Iacconi, M.~Fasiello, H.~Assadullahi, E.~Dimastrogiovanni, and D.~Wands, ``{Interferometer Constraints on the Inflationary Field Content},'' {\em JCAP}, vol.~03, p.~031, 2020.

\bibitem{Planck:2018jri}
Y.~Akrami {\em et~al.}, ``{Planck 2018 results. X. Constraints on inflation},'' {\em Astron. Astrophys.}, vol.~641, p.~A10, 2020.

\bibitem{Planck:2019kim}
Y.~Akrami {\em et~al.}, ``{Planck 2018 results. IX. Constraints on primordial non-Gaussianity},'' {\em Astron. Astrophys.}, vol.~641, p.~A9, 2020.

\bibitem{Andriot:2018mav}
D.~Andriot and C.~Roupec, ``Further refining the de sitter swampland conjecture,'' {\em Fortschritte der Physik}, vol.~67, p.~1800105, 02 2019.

\bibitem{Brennan:2017rbf}
T.~D. Brennan, F.~Carta, and C.~Vafa, ``{The String Landscape, the Swampland, and the Missing Corner},'' {\em PoS}, vol.~TASI2017, p.~015, 2017.

\bibitem{Obied:2018sgi}
G.~{Obied}, H.~{Ooguri}, L.~{Spodyneiko}, and C.~{Vafa}, ``{De Sitter Space and the Swampland},'' {\em arXiv e-prints}, p.~arXiv:1806.08362, June 2018.

\bibitem{Garg:2018reu}
S.~K. Garg and C.~Krishnan, ``{Bounds on Slow Roll and the de Sitter Swampland},'' {\em JHEP}, vol.~11, p.~075, 2019.

\bibitem{kinney2019}
W.~H. Kinney, S.~Vagnozzi, and L.~Visinelli, ``The zoo plot meets the swampland: mutual (in)consistency of single-field inflation, string conjectures, and cosmological data,'' {\em Classical and Quantum Gravity}, vol.~36, p.~117001, may 2019.

\bibitem{Kallosh:2019axr}
R.~Kallosh, A.~Linde, E.~McDonough, and M.~Scalisi, ``{dS Vacua and the Swampland},'' {\em JHEP}, vol.~03, p.~134, 2019.

\bibitem{Himmetoglu:2009qi}
B.~Himmetoglu, C.~R. Contaldi, and M.~Peloso, ``{Ghost instabilities of cosmological models with vector fields nonminimally coupled to the curvature},'' {\em Phys. Rev. D}, vol.~80, p.~123530, 2009.

\bibitem{BeltranJimenez:2017cbn}
J.~Beltran~Jimenez, L.~Heisenberg, R.~Kase, R.~Namba, and S.~Tsujikawa, ``{Instabilities in Horndeski Yang-Mills inflation},'' {\em Phys. Rev. D}, vol.~95, no.~6, p.~063533, 2017.

\bibitem{Golovnev:2011yc}
A.~Golovnev, ``{On cosmic inflation in vector field theories},'' {\em Class. Quant. Grav.}, vol.~28, p.~245018, 2011.

\bibitem{Gorji:2020vnh}
M.~A. Gorji, S.~A. Hosseini~Mansoori, and H.~Firouzjahi, ``{Inflation with multiple vector fields and non-Gaussianities},'' {\em JCAP}, vol.~11, p.~041, 2020.

\bibitem{Murata:2021vnb}
T.~Murata and T.~Kobayashi, ``{Dynamics of inflation with mutually orthogonal vector fields in a closed universe},'' {\em Phys. Rev. D}, vol.~104, no.~8, p.~083514, 2021.

\bibitem{jiro2009}
M.-a. Watanabe, S.~Kanno, and J.~Soda, ``Inflationary universe with anisotropic hair,'' {\em Phys. Rev. Lett.}, vol.~102, p.~191302, May 2009.

\bibitem{jiro2010a}
S.~Kanno, J.~Soda, and M.~aki Watanabe, ``Anisotropic power-law inflation,'' {\em Journal of Cosmology and Astroparticle Physics}, vol.~2010, pp.~024--024, dec 2010.

\bibitem{Prokopec:2005fb}
T.~Prokopec and W.~Valkenburg, ``{The Cosmology of the nonsymmetric theory of gravitation},'' {\em Phys. Lett. B}, vol.~636, pp.~1--4, 2006.

\bibitem{Koivisto:2009sd}
T.~S. Koivisto, D.~F. Mota, and C.~Pitrou, ``{Inflation from N-Forms and its stability},'' {\em JHEP}, vol.~09, p.~092, 2009.

\bibitem{Obata:2018ilf}
I.~Obata and T.~Fujita, ``{Footprint of Two-Form Field: Statistical Anisotropy in Primordial Gravitational Waves},'' {\em Phys. Rev. D}, vol.~99, no.~2, p.~023513, 2019.

\bibitem{Elizalde:2018rmz}
E.~Elizalde, S.~D. Odintsov, T.~Paul, and D.~S\'aez-Chill\'on~G\'omez, ``{Inflationary universe in $F(R)$ gravity with antisymmetric tensor fields and their suppression during its evolution},'' {\em Phys. Rev. D}, vol.~99, no.~6, p.~063506, 2019.

\bibitem{Rohm:1986}
R.~Rohm and E.~Witten, ``The antisymmetric tensor field in superstring theory,'' {\em Annals of Physics}, vol.~170, no.~2, pp.~454--489, 1986.

\bibitem{Ghezelbash:2009gf}
A.~M. Ghezelbash, ``{Kerr/CFT Correspondence in Low Energy Limit of Heterotic String Theory},'' {\em JHEP}, vol.~08, p.~045, 2009.

\bibitem{jiro2010b}
M.-a. Watanabe, S.~Kanno, and J.~Soda, ``{The Nature of Primordial Fluctuations from Anisotropic Inflation},'' {\em Progress of Theoretical Physics}, vol.~123, pp.~1041--1068, 06 2010.

\bibitem{jiro2013}
J.~Ohashi, J.~Soda, and S.~Tsujikawa, ``Anisotropic non-gaussianity from a two-form field,'' {\em Phys. Rev. D}, vol.~87, p.~083520, Apr 2013.

\bibitem{jiro2015}
A.~Ito and J.~Soda, ``Designing anisotropic inflation with form fields,'' {\em Phys. Rev. D}, vol.~92, p.~123533, Dec 2015.

\bibitem{2012JCAP...12..016M}
D.~J. {Mulryne}, J.~{Noller}, and N.~J. {Nunes}, ``{Three-form inflation and non-Gaussianity},'' {\em JCAP}, vol.~2012, p.~016, Dec. 2012.

\bibitem{DeFelice:2012jt}
A.~De~Felice, K.~Karwan, and P.~Wongjun, ``{Stability of the 3-form field during inflation},'' {\em Phys. Rev. D}, vol.~85, p.~123545, 2012.

\bibitem{Germani:2009iq}
C.~Germani and A.~Kehagias, ``{P-nflation: generating cosmic Inflation with p-forms},'' {\em JCAP}, vol.~03, p.~028, 2009.

\bibitem{Koivisto:2009ew}
T.~S. Koivisto and N.~J. Nunes, ``{Three-form cosmology},'' {\em Phys. Lett. B}, vol.~685, pp.~105--109, 2010.

\bibitem{Ajith:2022wia}
A.~Ajith and S.~Panda, ``{Inflation using a triplet of antisymmetric tensor fields},'' {\em Eur. Phys. J. C}, vol.~83, no.~8, p.~770, 2023.

\bibitem{Aashish:2018lhv}
S.~Aashish, A.~Padhy, S.~Panda, and A.~Rana, ``{Inflation with an antisymmetric tensor field},'' {\em Eur. Phys. J. C}, vol.~78, no.~11, p.~887, 2018.

\bibitem{Aashish:2019zsy}
S.~Aashish, A.~Padhy, and S.~Panda, ``{Avoiding instabilities in antisymmetric tensor field driven inflation},'' {\em Eur. Phys. J. C}, vol.~79, no.~9, p.~784, 2019.

\bibitem{Aashish:2020mlw}
S.~Aashish, A.~Padhy, and S.~Panda, ``{Gravitational waves from inflation with antisymmetric tensor field},'' {\em JCAP}, vol.~12, pp.~004--004, 2020.

\bibitem{Aashish:2021gdf}
S.~Aashish, A.~Ajith, S.~Panda, and R.~Thakur, ``{Inflation with antisymmetric tensor field: new candidates},'' {\em JCAP}, vol.~04, no.~04, p.~043, 2022.

\bibitem{Dodelson:2003ft}
S.~Dodelson, {\em {Modern Cosmology}}.
\newblock Academic Press, Elsevier Science, 2003.

\bibitem{Guzzetti:2016mkm}
M.~C. Guzzetti, N.~Bartolo, M.~Liguori, and S.~Matarrese, ``{Gravitational waves from inflation},'' {\em Riv. Nuovo Cim.}, vol.~39, no.~9, pp.~399--495, 2016.

\bibitem{Altschul:2009ae}
B.~Altschul, Q.~G. Bailey, and V.~A. Kostelecky, ``{Lorentz violation with an antisymmetric tensor},'' {\em Phys. Rev. D}, vol.~81, p.~065028, 2010.

\bibitem{Odintsov:2019clh}
S.~D. Odintsov and V.~K. Oikonomou, ``{Inflationary Phenomenology of Einstein Gauss-Bonnet Gravity Compatible with GW170817},'' {\em Phys. Lett. B}, vol.~797, p.~134874, 2019.

\bibitem{Weinberg:2008zzc}
S.~Weinberg, {\em {Cosmology}}.
\newblock 2008.

\bibitem{Ma:1995ey}
C.-P. Ma and E.~Bertschinger, ``{Cosmological perturbation theory in the synchronous and conformal Newtonian gauges},'' {\em Astrophys. J.}, vol.~455, pp.~7--25, 1995.

\bibitem{Gumrukcuoglu:2016jbh}
A.~E. G\"umr\"uk\c{c}\"uo\u{g}lu, S.~Mukohyama, and T.~P. Sotiriou, ``{Low energy ghosts and the Jeans\textquoteright{} instability},'' {\em Phys. Rev. D}, vol.~94, no.~6, p.~064001, 2016.

\bibitem{Kundu:2011sg}
S.~Kundu, ``{Inflation with General Initial Conditions for Scalar Perturbations},'' {\em JCAP}, vol.~02, p.~005, 2012.

\bibitem{Naoz:2013wla}
S.~Naoz and R.~Narayan, ``{Generation of Primordial Magnetic Fields on Linear Over-density Scales},'' {\em Phys. Rev. Lett.}, vol.~111, p.~051303, 2013.

\bibitem{Demozzi:2009fu}
V.~Demozzi, V.~Mukhanov, and H.~Rubinstein, ``{Magnetic fields from inflation?},'' {\em JCAP}, vol.~08, p.~025, 2009.

\bibitem{Basak:2014qea}
A.~Basak and S.~Shankaranarayanan, ``{Super-inflation and generation of first order vector perturbations in ELKO},'' {\em JCAP}, vol.~05, p.~034, 2015.

\end{thebibliography}
\newpage
\appendix
\section{Differential equations for the vector modes}\label{app_vec}
\begin{flushleft}
    1. \textbf{Subhorizon limit}
\end{flushleft}
\begin{equation}
    \begin{aligned}
        \frac{dC_x^{\phi}}{dp} &= \alpha_c, \hspace{3em} \frac{dC_y^{\phi}}{dp} = \beta_c, \hspace{3em} \frac{dU_x^i}{dp} = \alpha_u, \\
        \frac{dU_y^i}{dp} &= \beta_u, \hspace{3em} \frac{dV_x}{dp} = \alpha_v  , \hspace{3em} \frac{dV_y}{dp} = \beta_v \\
        \frac{d\alpha_c}{dp} &= - \frac{C^{\phi}_x}{3} - \frac{C^{\phi}_y}{3} - \frac{\alpha_v}{3 k \tau a^2} - \frac{2 \beta_v}{3 k \tau a^2} - \frac{2 U^i_x}{3 k \tau a} + \frac{U^i_y}{3 k \tau a} - \frac{2 H V_x}{3 k^2 a} - \frac{4 H V_y}{3 k^2 a}
 \\
\frac{d\beta_c}{dp} &= - \frac{C^{\phi}_x}{3} - \frac{C^{\phi}_y}{3} 
+ \frac{2 \alpha_v}{3 k \tau a^2} + \frac{\beta_v}{3 k \tau a^2} 
+ \frac{U^i_x}{3 k \tau a} - \frac{2 U^i_y}{3 k \tau a} 
+ \frac{4 H V_x}{3 k^2 a} + \frac{2 H V_y}{3 k^2 a}
 \\
\frac{d \alpha_u}{dp} &= \frac{1}{3ka} \left( C^{\phi}_x + C^{\phi}_y \right)
+ \frac{\alpha_v}{3\tau a}
+ \frac{\beta_v}{a}
+ \frac{5\beta_v}{3\tau a}
- \frac{a\,\alpha_u\,H}{k}
+ \frac{5 U^i_x}{3\tau}
- \frac{U^i_y}{3\tau}
+ \frac{2 H V_x}{3k}
+ \frac{10 H V_y}{3k}
 \\
\frac{d\beta_u}{dp} &= \frac{1}{3ka} \left( C^{\phi}_x + C^{\phi}_y \right)
- \frac{\alpha_v}{a}
- \frac{5\alpha_v}{3\tau a}
- \frac{\beta_v}{3\tau a}
- \frac{a\,\beta_u\,H}{k}
- \frac{U^i_x}{3\tau}
+ \frac{5 U^i_y}{3\tau}
- \frac{10 H V_x}{3k}
- \frac{2 H V_y}{3k}  \\
\frac{d \alpha_v}{dp} &= k \tau a^2 \beta_c + a\,\beta_u + \tau a\,\beta_u + \tau V_x
\\
\frac{d\beta_v}{dp} &=- k \tau a^2 \alpha_c - a\,\alpha_u - \tau a\,\alpha_u + \tau V_y
 \\
    \end{aligned}
\end{equation}
where $C^{\phi}_j\equiv \phi \ C_j$ and $U^i_j\equiv i\ U_i$
\begin{flushleft}
    2. \textbf{Complete expression}
\end{flushleft}
\begin{equation}
    \begin{aligned}
       \frac{dC_x^{\phi}}{dp} &= \alpha_c, \hspace{3em} \frac{dC_y^{\phi}}{dp} = \beta_c, \hspace{3em} \frac{dU_x^i}{dp} = \alpha_u, \\
        \frac{dU_y^i}{dp} &= \beta_u, \hspace{3em} \frac{dV_x}{dp} = \alpha_v  , \hspace{3em} \frac{dV_y}{dp} = \beta_v \\
        \frac{d\alpha_c}{dp} &= - \frac{C^{\phi}_x}{3}
- \frac{16 C^{\phi}_x}{3 p^2}
- \frac{C^{\phi}_y}{3}
+ \frac{2 C^{\phi}_y}{3 p^2}
+ \frac{10 \alpha_c}{3 p}
+ \frac{4 H \alpha_u}{3 k^2}
- \frac{H^2 p^2 \alpha_v}{3 k^3 \tau}
- \frac{2 \beta_c}{3 p}
- \frac{2 H \beta_u}{3 k^2}
- \frac{2 H^2 p^2 \beta_v}{3 k^3 \tau} \\
& \quad - \frac{8 H U^i_x}{3 k^2 p}
- \frac{2 H p U^i_x}{3 k^2 \tau}
+ \frac{4 H U^i_y}{3 k^2 p}
+ \frac{H p U^i_y}{3 k^2 \tau}
- \frac{2 H^2 p V_x}{3 k^3}
- \frac{4 H^2 p V_y}{3 k^3} \\
     \frac{d\beta_c}{dp} &= - \frac{C^{\phi}_x}{3}
+ \frac{2 C^{\phi}_x}{3 p^2}
- \frac{C^{\phi}_y}{3}
- \frac{16 C^{\phi}_y}{3 p^2}
- \frac{2 \alpha_c}{3 p}
- \frac{2 H \alpha_u}{3 k^2}
+ \frac{2 H^2 p^2 \alpha_v}{3 k^3 \tau}
+ \frac{10 \beta_c}{3 p}
+ \frac{4 H \beta_u}{3 k^2}
+ \frac{H^2 p^2 \beta_v}{3 k^3 \tau}
\\
&\quad + \frac{4 H U^i_x}{3 k^2 p}
+ \frac{H p U^i_x}{3 k^2 \tau}
- \frac{8 H U^i_y}{3 k^2 p}
- \frac{2 H p U^i_y}{3 k^2 \tau}
+ \frac{4 H^2 p V_x}{3 k^3}
+ \frac{2 H^2 p V_y}{3 k^3} \\
\frac{d\alpha_u}{dp} &= \frac{22 k^2 C^{\phi}_x}{3 H p^3}
+ \frac{k^2 C^{\phi}_x}{3 H p}
- \frac{2 k^2 C^{\phi}_y}{3 H p^3}
+ \frac{k^2 C^{\phi}_y}{3 H p}
- \frac{10 k^2 \alpha_c}{3 H p^2}
- \frac{4 \alpha_u}{3 p}
+ \frac{H p \alpha_v}{3 k \tau}
+ \frac{2 k^2 \beta_c}{3 H p^2}
+ \frac{2 \beta_u}{3 p}
+ \frac{H p \beta_v}{k}
\\
&\quad + \frac{5 H p \beta_v}{3 k \tau}
+ \frac{20 U^i_x}{3 p^2}
+ \frac{5 U^i_x}{3 \tau}
- \frac{4 U^i_y}{3 p^2}
- \frac{U^i_y}{3 \tau}
+ \frac{2 H V_x}{3 k}
+ \frac{10 H V_y}{3 k}\\
\frac{d\beta_u}{dp} &=  - \frac{2 k^2 C^{\phi}_x}{3 H p^3}
+ \frac{k^2 C^{\phi}_x}{3 H p}
+ \frac{22 k^2 C^{\phi}_y}{3 H p^3}
+ \frac{k^2 C^{\phi}_y}{3 H p}
+ \frac{2 k^2 \alpha_c}{3 H p^2}
+ \frac{2 \alpha_u}{3 p}
- \frac{H p \alpha_v}{k}
- \frac{5 H p \alpha_v}{3 k \tau}
- \frac{10 k^2 \beta_c}{3 H p^2}
- \\
&\quad \frac{4 \beta_u}{3 p}
- \frac{H p \beta_v}{3 k \tau}
- \frac{4 U^i_x}{3 p^2}
- \frac{U^i_x}{3 \tau}
+ \frac{20 U^i_y}{3 p^2}
+ \frac{5 U^i_y}{3 \tau}
- \frac{10 H V_x}{3 k}
- \frac{2 H V_y}{3 k} \\
\frac{d\alpha_u}{dp} &= - \frac{2 \alpha_v}{p}
+ \frac{k^3 \tau \beta_c}{H^2 p^2}
+ \frac{k \beta_u}{H p}
+ \frac{k \tau \beta_u}{H p}
+ \frac{k U^i_y}{H p^2}
- \frac{k \tau U^i_y}{H p^2}
+ \tau V_x
- \frac{4 \tau V_x}{p^2} \\
\frac{d\beta_u}{dp} &= - \frac{k^3 \tau \alpha_c}{H^2 p^2}
- \frac{k \alpha_u}{H p}
- \frac{k \tau \alpha_u}{H p}
- \frac{2 \beta_v}{p}
- \frac{k U^i_x}{H p^2}
+ \frac{k \tau U^i_x}{H p^2}
+ \tau V_y
- \frac{4 \tau V_y}{p^2}
     \end{aligned}
\end{equation}
\end{document}